\documentclass{aa}

\usepackage{natbib}
\usepackage{graphicx}
\usepackage{subfigure}
\usepackage{amssymb}
\usepackage[dvipdfm,colorlinks=true,linkcolor=blue,citecolor=blue,urlcolor=blue]{hyperref} 
\newcommand\T{\rule{0pt}{2.1ex}}
\newcommand\B{\rule[-1.7ex]{0pt}{0pt}}
\def\Ha{H$\alpha$}
\def\Hb{H$\beta$}
\def\Hg{H$\gamma$}
\def\Hd{H$\delta$}
\def\He{H$\epsilon$}
\def\Oii{{\sc [Oii]}}

\def\Oiii{{\sc [Oiii]}}

\def\wings{{\sc wings}}
\def\sdss{{\sc sdss}}
\def\ea{{\it e(a)}}
\def\eb{{\it e(b)}}
\def\ec{{\it e(c)}}
\def\k{{\it k}}
\def\ka{{\it k+a}}
\def\ak{{\it a+k}}

\begin{document}

\title{WINGS-SPE III:\\
Equivalent width measurements, spectral properties and evolution of local cluster galaxies.  
\thanks{Based on observations taken at the Anglo Australian Telescope (3.9 m- AAT), and at the William Herschel Telescope (4.2 m- WHT)}
\thanks{Table \ref{tab:cat} is available in electronic form both at the CDS via anonymous ftp to cdsarc.u-strasbg.fr (130.79.128.5)
or via http://cdsweb.u-strasbg.fr/cgi-bin/qcat?J/A+A/????/?????, and by querying the \wings \ database at http://web.oapd.inaf.it/wings/new/index.html}}

\author{J. Fritz\inst{1} \and B.~M. Poggianti\inst{2} \and A. Cava\inst{3} \and A. Moretti\inst{4,2} \and J. Varela\inst{5} \and D. Bettoni\inst{2} \and W.~J. Couch\inst{6,7} \and M. D'Onofrio\inst{4} \and A.~Dressler\inst{8} \and G. Fasano\inst{2} \and P. Kj\ae rgaard\inst{9} \and P. Marziani\inst{2} \and M. Moles\inst{5} \and A. Omizzolo\inst{2,10}.}
\offprints{Jacopo Fritz,\\ 
           \email{jacopo.fritz@UGent.be}}
\institute{Sterrenkundig Observatorium Vakgroep Fysica en Sterrenkunde Universiteit Gent, Krijgslaan 281, S9  9000 Gent\\
\email{jacopo.fritz@UGent.be}
\and
INAF-Osservatorio Astronomico di Padova, vicolo Osservatorio 5, 35122 Padova, Italy
\and 
Observatoire de Gen{\`e}ve, Universit{\'e} de Gen{\`e}ve, 51 Ch. des Maillettes, 1290 Versoix, Switzerland
\and
 Dipartimento di Astronomia, vicolo Osservatorio 2, 35122 Padova, Italy
\and
 Centro de Estudios de F\'\i sica del Cosmos de Arag\'on (CEFCA), Plaza San Juan 1, planta 2, E-44001 Teruel, Spain
\and
 Centre for Astrophysics and Supercomputing, Swinburne University of Technology, Melbourne Australia
\and
Australian Astronomical Observatory, Sydney Australia
\and
 Observatories of the Carnegie Institution of Washington, Pasadena, CA 91101, USA
\and
 Niels Bohr Institute, Juliane Maries Vej 30, 2100 Copenhagen, Denmark
\and
Specola Vaticana, 00120 Stato Citt\`a del Vaticano}

\date{Received ...; accepted ...}

\titlerunning{EW of \wings \ galaxies}
\authorrunning{Fritz J. et al.}

\abstract 
{Cluster galaxies are the ideal sites to look at when studying the influence of the environment on the various aspects of galaxies' evolution, such as the changes in their stellar content and morphological transformations. In the framework of \wings, the WIde-field Nearby Galaxy-cluster Survey, we have obtained optical spectra for $\sim 6000$ galaxies selected in fields centered on 48 local ($0.04< z <0.07$) X-ray selected clusters to tackle these issues. }
{By classifying the spectra based on given spectral lines, we investigate the frequency of the various spectral types as a function both of the clusters' properties and of the galaxies' characteristics. In this way, using the same classification criteria adopted for studies at higher redshift, we can consistently compare the properties of the local cluster population to those of their more distant counterparts.}
{We describe a method we have developed to automatically measure the equivalent width of spectral lines in a robust way even in spectra with a non optimal signal to noise. Like this, we can derive a spectral classification  reflecting the stellar content, based on the presence and strength of the \Oii \ and \Hd \ lines.}
{After a quality check, we are able to measure 4381 of the $\sim 6000$ originally observed spectra, in the fields of 48 clusters, 2744 of which are spectroscopically confirmed cluster members. The spectral classification is then analyzed as a function of galaxies' luminosity, stellar mass, morphology, local density and host cluster's global properties, and compared to higher redshift samples (MORPHS and EDisCS). The vast majority of galaxies in the local clusters population are passive objects, being also the most luminous and massive. At a magnitude limit of M$_V<-18$, galaxies in a post--starburst phase represent only $\sim 11$\% of the cluster population and this fraction is reduced to $\sim 5$\% at  M$_V<-19.5$, which compares to the 18\% at the same magnitude limit for high-z clusters. ``Normal'' star forming galaxies (\ec) are proportionally more common in local clusters.}
{The relative occurrence of post--starbursts suggests a very similar quenching efficiency in clusters at redshifts in the $0$ to $\sim 1$ range. Furthermore, more important than the global environment, the local density seems to be the main driver of galaxy evolution in local clusters, at least with respect to their stellar populations content.}


\maketitle


\section{Introduction}
It is now widely accepted that the evolution of galaxies' properties
such as morphology, gas and dust content and stellar population 
properties, has
a very strong dependence on the environment where galaxies are found.
Galaxies we observe in local clusters and in the field not only present different
characteristics, but also the way that led them to be as we observe
them now, that is their evolution, is different. By evolution, in this context, we mean
the transformations that galaxies undergo both from the morphological
aspect and with respect to their stellar population content, as
a function of the cosmic time.

A striking example of this environmental dependency is the paucity of blue, 
late-type galaxies at the centre of local clusters, at odd with what it was found
in higher redshift rich clusters where a large population of blue, 
star-forming galaxies was already discovered more than 3 decades ago
 \citep{butcher78a,butcher78b,butcher84}. 
This is one of the most clear examples of the consequences of 
the evolutionary effects observable in
galaxies in dense environments, and evidences for it have been gathered during
the following years \citep[see, among the others,][]{couch94,lavery97,margoniner00,
depropris03,depropris04,haines09}. 

Since then, galaxy clusters have proved to be a very important place to 
study galactic evolution, and were used to probe an extremely wide 
range of physical conditions, from the dense cores, to the outermost, 
low-density regions. This was achieved, during the past years, both 
from a morphological perspective \citep[see, e.g.,][]{hausman78,dressler80,
buote95,dressler97,fasano00,vandokkum01,goto04,sanchezblazques09,
poggianti09} and from the point of view of the stellar populations hosted 
by these galaxies \citep[e.g.][]{couch87,dressler83,dressler92,dressler99,
postman01,dressler13,gladders13}.

Furthermore, evidences are already in place since many years for the fact 
that the cluster environment acts not only on a ``global'' scale, but also
``locally''. \cite{prugniel99}, e.g., demonstrated how the presence of young
stellar populations is affected by the local environment, showing that 
galaxies hosting recent star formation tend to prefer the less dense environment
of the field or of poor groups. On the other side, the emission--line galaxies 
fraction was found to decrease towards the centre of dense clusters
\citep[e.g.][]{biviano97,dale01}, proving the importance of the local 
conditions. Similarly, the projected distance from the cluster's centre was
found to be the main factor driving the differences between passive ellipticals 
and actively star forming spirals \citep[see][and references therein]{thomas06},
proving the importance of the global environment. 

The best approach when trying to understand how blue, actively star forming
galaxies so common in high-redshift galaxy clusters, turn into red and 
passive systems, is probably to focus on those
objects which are experiencing a transition in between the two phases.
Among these, there 
are the so called E+A galaxies \citep{dressler83,zabludoff96}, also named \ka \ and \ak \ 
\citep{dressler99}, or simply ``post--starburst''. Their designation 
is derived from the characteristics in their spectra, which lack lines in emission 
and present strong Balmer absorption lines,
typical of A-class stars (hence the ``A'' in their names), having a lifetime of some
hundreds of Myr, up to about 1 Gyr. For this reason, such a spectral signature 
is very often used to identify a burst or an episode of enhanced star formation 
on those time scales. 

The presence of strong hydrogen lines in absorption,  
and the concomitant absence of any emission line in the optical, can be 
easily achieved as the duration of the two phenomena producing them is
different, with emission lines being typically produced from the intense
ionizing continuum of very young (i.e. 
ages less than $\sim 2\times 10^7$ years) stars. Such a combination
can appear as the result of a star bursting episode being observed 
few tens of Myr after the star formation has stopped \citep[e.g.][]{poggianti99}.
Alternatively, a truncation in an otherwise continuous, regularly star 
forming pattern, might produce moderately strong Balmer absorption 
lines as well \citep[e.g.][]{newberry90,leonardi96}.

Hence, in most cases, these are galaxies which have recently
experienced a quenching of the star formation activity and are on their way to 
become passively evolving objects. Thus, despite the fact that they are quite 
rare if compared to the whole galactic population 
\citep[see][and references therein]{quintero04}, they may be one of the
important keys to address galaxy evolution questions. 

E+A galaxies were initially believed to be a prerogative of galaxy 
clusters, and a number of mechanisms have been proposed to explain their
formation, including high speed galaxy--galaxy interactions \citep[e.g.][]{zabludoff96}, 
mergers \citep{bekki01}, gas stripping by ram--pressure from the hot intracluster
gas \citep{tonnesen07,ma08}. The latter in particular, detected in X--rays, 
is believed to be one of the main actors in a galaxy's life, causing the 
quenching of the star formation in those systems which are accreting 
into clusters, through the efficient removal of their gas reservoirs.

Optical spectra contain the information we need in order 
to study, through the analysis of their integrated light, the stellar 
content of a galaxy: the stellar mass, the ages of stellar populations, 
their metallicity and the dust content are all characteristics that can, 
at least to some degree, be derived from such kind of data.

The \wings\footnote[1]{You can refer to the \wings \ website for a
description of the project and updated details on the survey and its
products: \tt{http://web.oapd.inaf.it/wings/new/index.html}} project 
\citep{fasano06}, and its spectroscopic follow--up \citep{cava09},
is providing the largest set of homogeneous spectroscopic data for
galaxies belonging to nearby clusters. Spectral lines are of
fundamental importance to characterize the stellar content and
properties of a galaxy, providing also a quick and simple way to
broadly classify them, according to their star formation history. 
The reader can refer to \cite{poggianti99} for an interpretation,
from a theoretical point of view, of the meaning of the various 
spectroscopic classes. 

In this paper we will describe the details of a 
method used to automatically measure the equivalent width 
(EW) of the most 
prominent spectral lines in \wings \ optical spectra. The method,
presented in \cite{fritz07} for the first time, is able to deal with spectra 
of low 
signal-to-noise, and provides a robust estimate of the uncertainty
as well. We present and describe the catalogue of rest-frame 
EW, providing a spectral classification, and we compare the 
spectral properties of $\sim 2000$ galaxies to other characteristics
such as their luminosity, mass, morphology and to the environment
where they reside.

The paper outline is as follows: we briefly present the dataset in
Sect.~\ref{sec:data} and explain and discuss the measurement method
in Sect.~\ref{sec:method}. In Sect.~\ref{sec:spec} we point out
some observational issues and describe the spectroscopic
classification criteria, while in Sect.~\ref{sec:properties} we
describe the properties of galaxies in our sample, based on their
spectroscopic classification in relation with both other galaxies'
properties (such as luminosity, stellar mass, etc.) and with cluster's
properties as well. This is followed by a comparison with high 
redshift cluster galaxies (Sect.~\ref{sec:highz}), and by a 
summary of our results (Sect.~\ref{sec:conclusions}). 
Details on the EW catalog, together with
an example, are given in Appendix \ref{sec:catalog}.  

Throughout this paper we have adopted the usual convention of 
identifying emission lines with negative values of the EWs, and absorption 
lines with positive ones, with all the quantities given at
restframe. Whenever stellar masses are used, we always refer 
to a \cite{salpeter55} IMF, with masses in the $0.15-120$ 
M$_\odot$ range, and stellar mass definition n.2, as discussed in
\cite{longhetti09}. Furthermore, we assume a standard $\Lambda$ cold dark
matter ($\Lambda$CDM) cosmology, with $H_0=70$, $\Omega_m=0.3$ and
$\Omega_\Lambda=0.7$. 

\section{The dataset}\label{sec:data}
Out of the 77 cluster fields imaged by the \wings \ photometric survey
\citep[see][for the presentation of the photometric catalog]{varela09}, 
48 were also followed-up spectroscopically. While the
reader should refer to \cite{cava09} for a complete description of the
spectroscopic sample, including data reduction, quality check, and
completeness analysis, here we will briefly summarize the features
that are more relevant for this work's purposes.
The results presented in this paper are based on \wings \ spectra and
catalogs discussed in \cite{cava09}. Our apparent magnitude limit ($V
\sim 20$) is 1.5 to 2.0 mag deeper than the 2dFRS and \sdss \ surveys,
respectively, and this is, in general, reflected in a higher mean
number of member galaxies detected per cluster.

The target selection was based on the available \wings \ optical B and V
photometry \citep{varela09}, adopting a generous red cut well above the
cluster red sequence. The aim of the selection strategy was
to maximize the chances of observing galaxies at the cluster redshift
without biasing the sample. 

A good knowledge of the completeness level of the spectroscopic
observations is required by the analysis that we will present, as this
is a factor that must be accounted for in the derivation of luminosity
functions, as well as anytime we want to use the
spectroscopic sample to study magnitude--dependent properties (e.g. the
different galaxy population fractions inside clusters, see
\citealt{poggianti06}). We have computed both magnitude and geometrical
(to account for the crowding of the fibers spectroscope near the clusters' centre) 
completeness which we describe in more detail in Appendix~\ref{sec:completeness}.

Medium resolution spectra for $\sim 6000$ galaxies were obtained
during several runs at the William Herschel Telescope (WHT) and at the
Anglo Australian Telescope (AAT) with multi-fiber spectrographs
(WYFFOS and 2dF, respectively), yielding reliable redshift
measurements. The fiber apertures were $1''.6$ and $2''$, and the
spectral resolution $\sim 6$ and $\sim 9$ \AA \ FWHM for WHT and AAT
spectra, respectively. The wavelength coverage ranges from $\sim 3590$
to $\sim 6800$ \AA \ for the WHT observations, while spectra taken at
the AAT cover the $\sim 3600$ to $\sim 8000$ \AA \ domain. Note also
that, for just one observing run at WHT (during which 3 clusters were
observed), due to a different setup, the spectral resolution was 
$\sim 3$ \AA \ FWHM, with the
spectral coverage ranging from $\sim 3600$ to $\sim 6890$ \AA.

The spectra were observed by adopting two configurations, depending on
the galaxies' flux: bright and faint \citep[see][for further details]{cava09}. 
In principle, an issue possibly affecting our observations, is the presence of 
saturated emission lines. While this would not be an issue for the spectral 
classification, it could indeed affect the lines' flux measurements. However,
saturation does not represent a relevant problem for our dataset, as spectra taken in the 
bright configuration are mostly luminous early-type galaxies, with no or at
most faint emission lines. Spectra from the faint configuration, on the 
other side, belong to galaxies which are not bright enough to display 
saturated emission lines.

\section{The method}\label{sec:method}
Given the rather large number of spectra in the \wings \ spectroscopic
sample, we developed an automatic method capable of yielding accurate
measurements of the most important spectral lines. In table
\ref{tab:lines} we list the 14 lines that are measured for
each spectrum.

\begin{table}
\centering
\begin{tabular}{l | r | c | c}
\B
{\sc line}         & $\lambda$ [\AA]  &  {\sc em/abs} &  {\sc criterion} \\ 
\hline
\hline
\T
\Ha+{\sc Nii}     	& $6563$ 		& {\sc e}     & {\sc fixed} $\Delta\lambda$ \\ 
\Ha     	         & $6563$ 		& {\sc a}     & {\sc $1^{st}$ slope change} \\ 
Na ({\sc d})      	& $5893$          & {\sc a}     & {\sc $1^{st}$ slope change} \\ 
Mg		       	& $5177$ 		& {\sc a}     & {\sc $1^{st}$ slope change} \\  
\Oiii                   	& $5007$ 		& {\sc a}     & {\sc $1^{st}$ slope change} \\ 
\Hb                    	& $4861$ 		& {\sc e}     & {\sc $1^{st}$ slope change} \\  
\Hb                    	& $4861$ 		& {\sc a}     & {\sc $1^{st}$ slope change} \\  
\Hb                    	& $4861$ 		& {\sc e+a} & {\sc $2^{nd}$ slope change}\\ 
\Hg                    	& $4341$ 		& {\sc e}     & {\sc $1^{st}$ slope change} \\ 
\Hg                    	& $4341$ 		& {\sc a}     & {\sc $1^{st}$ slope change} \\ 
\Hg                    	& $4341$ 		& {\sc e+a} & {\sc $2^{nd}$ slope change}\\ 
CO {\sc g}-band& $4305$		& {\sc a}     & {\sc $1^{st}$ slope change} \\ 
\Hd                    	& $4101$ 		& {\sc e}     & {\sc $1^{st}$ slope change} \\  
\Hd                    	& $4101$ 		& {\sc a}     & {\sc $1^{st}$ slope change} \\ 
\Hd                    	& $4101$ 		& {\sc e+a} & {\sc $2^{nd}$ slope change}\\ 
\He+Ca{\sc ii (h)} & $3969$ 	& {\sc e}     & {\sc $1^{st}$ slope change} \\  
\He+Ca{\sc ii (h)} & $3969$ 	& {\sc a}     & {\sc $1^{st}$ slope change} \\  
\He+Ca{\sc ii (h)} & $3969$ 	& {\sc e+a} & {\sc $2^{nd}$ slope change}\\ 
Ca{\sc ii (k)}     	& $3934$ 		& {\sc a}     & {\sc $1^{st}$ slope change} \\ 
H$\zeta$	   	& $3889$ 		& {\sc e}     & {\sc $1^{st}$ slope change} \\  
H$\zeta$	   	& $3889$ 		& {\sc a}     & {\sc $1^{st}$ slope change} \\  
H$\zeta$	   	& $3889$ 		& {\sc e+a} & {\sc $2^{nd}$ slope change}\\ 
H$\eta$ 	   	& $3835$ 		& {\sc e}     & {\sc $1^{st}$ slope change} \\ 
H$\theta$	   	& $3798$ 		& {\sc a}     & {\sc $1^{st}$ slope change} \\ 
\Oii               	& $3727$ 		& {\sc e}     & {\sc $1^{st}$ slope change} \\
\hline
\end{tabular}				    	
\normalsize
\caption{The list of lines, whose equivalent width was measured, is 
reported here, together with their respective theoretical central $\lambda$.
On the fourth column we report the adopted criteria for the determination 
of their underlying continuum (see details in Sect.~\ref{sec:method}), 
distinguishing, if needed, the occurrence of emission and/or absorption
pattern (third column).}
\label{tab:lines}
\end{table}

Some automatic measurement procedures are already available in the
literature (such as ARES, \citealt{sousa07}, DAOSPEC,
\citealt{stetson08}, PACCE, \citealt{riffel11} or, more recently, TAME
by \citealt{kang_lee12}). While in general they are mainly focused on
the measurement of absorption lines, and in spectra with a high S/N,
we would ideally like to be able to use the information from spectra 
where the lines are still detectable even though somehow affected by the
noise. For example, as \cite{goto03} noted, an EW value derived by 
means of a Gaussian fit to the line, yields reliable results only in high
S/N spectra, unless the fit is performed interactively.
Furthermore, we are also interested in emission lines and in
the ``mixed'' case, in which absorption and emission profiles overlap.

When measuring an EW in a spectrum, the choice of the continuum turns out
to be a critical issue. In general, one would require that the line
profile is fully included, and that the measurement is made without
being affected by other lines in the proximity, or by noisy
features such as spikes, which could dramatically change the continuum
level, and hence the EW value. Defining the line profile by adopting a fixed wavelength
range, centered on the theoretical wavelength characterizing each line, can
be an unreliable choice because: 1) the line width can depend on the
EW itself (being larger as its value increases) or on the galactic
kinematic and 2) on a blind-measure approach, it is not possible to
verify whether the noise is dominating both the line and the continuum 
level. The method we adopted for the lines measurements has already 
been described in \cite{fritz07}, where we have preliminarily used it to 
constrain the lines' intensity on a theoretical spectral model. Here we 
will recall the most salient features and go into a more detailed description.

\begin{figure}[!t]
\centering
\includegraphics[height=.5\textwidth]{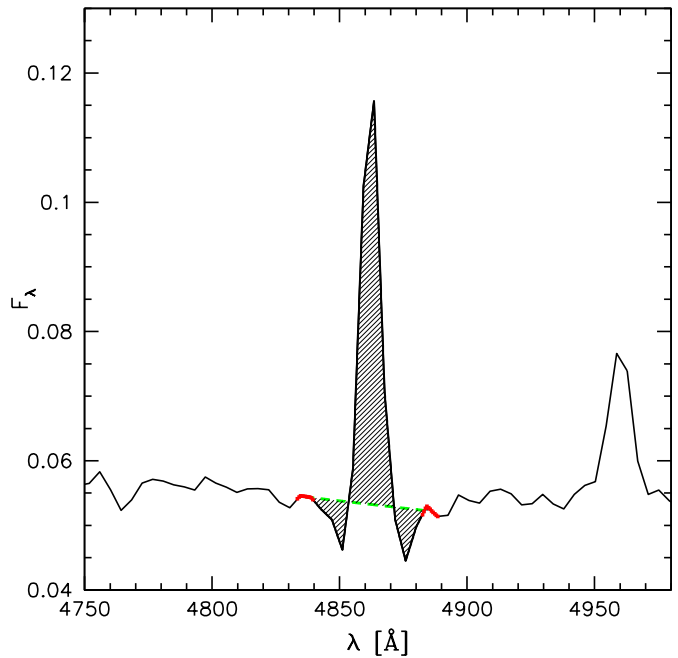}
\caption{Here we show how we define the areas that are used to compute
an equivalent width. In this example we plot an \Hb \ profile which 
includes both an absorption and an emission component. The green-dashed
line is meant to approximate the spectral continuum while its length,
$\Delta \lambda$, is used also to define the wavelength range
over which the line itself is measured.}
\label{fig:ewex}
\end{figure}

The algorithm measures the equivalent width of a line, no matter if in
absorption, emission, or including both components, by choosing the
appropriate value of the Full Width at Zero Intensity 
 --$\Delta \lambda$-- which is symmetric with respect to
the nominal (theoretical) line's centre (see table
\ref{tab:lines}). As we show in Fig. \ref{fig:ewex}, the edges of this
interval are used to define both the spectral continuum (red markers
at the extremes of the green-dashed line) and the range over which the
line is measured. The spectral continuum is approximated by a straight
line --green-dashed in Fig \ref{fig:ewex}-- and the EW value is found
by summing the dashed areas, divided by the average continuum
value.
 
\begin{figure*} 
\centering
\begin{tabular}{rr}
\rotatebox{270}{\includegraphics[height=.75\textwidth]{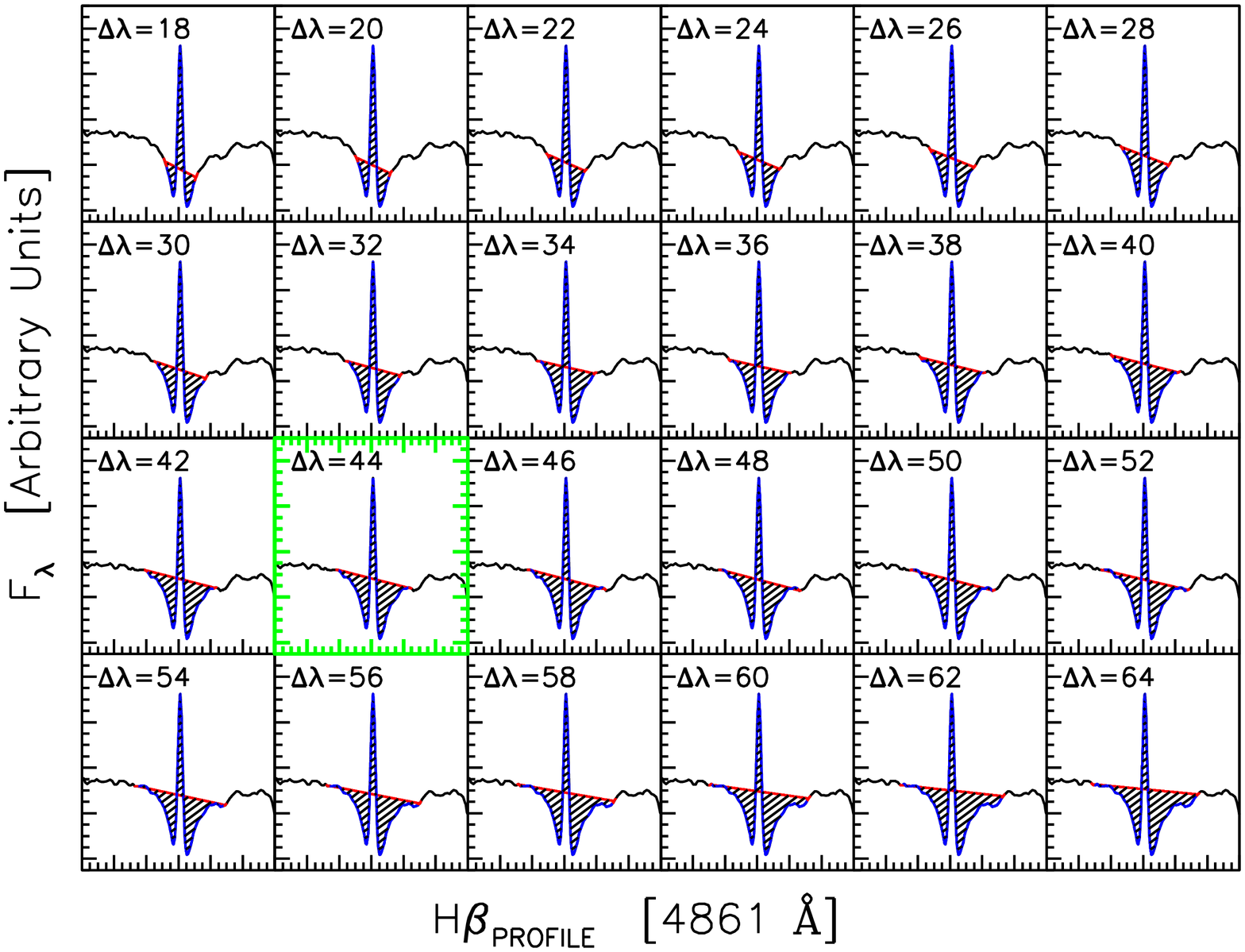}}\\
\rotatebox{270}{\includegraphics[height=.75\textwidth]{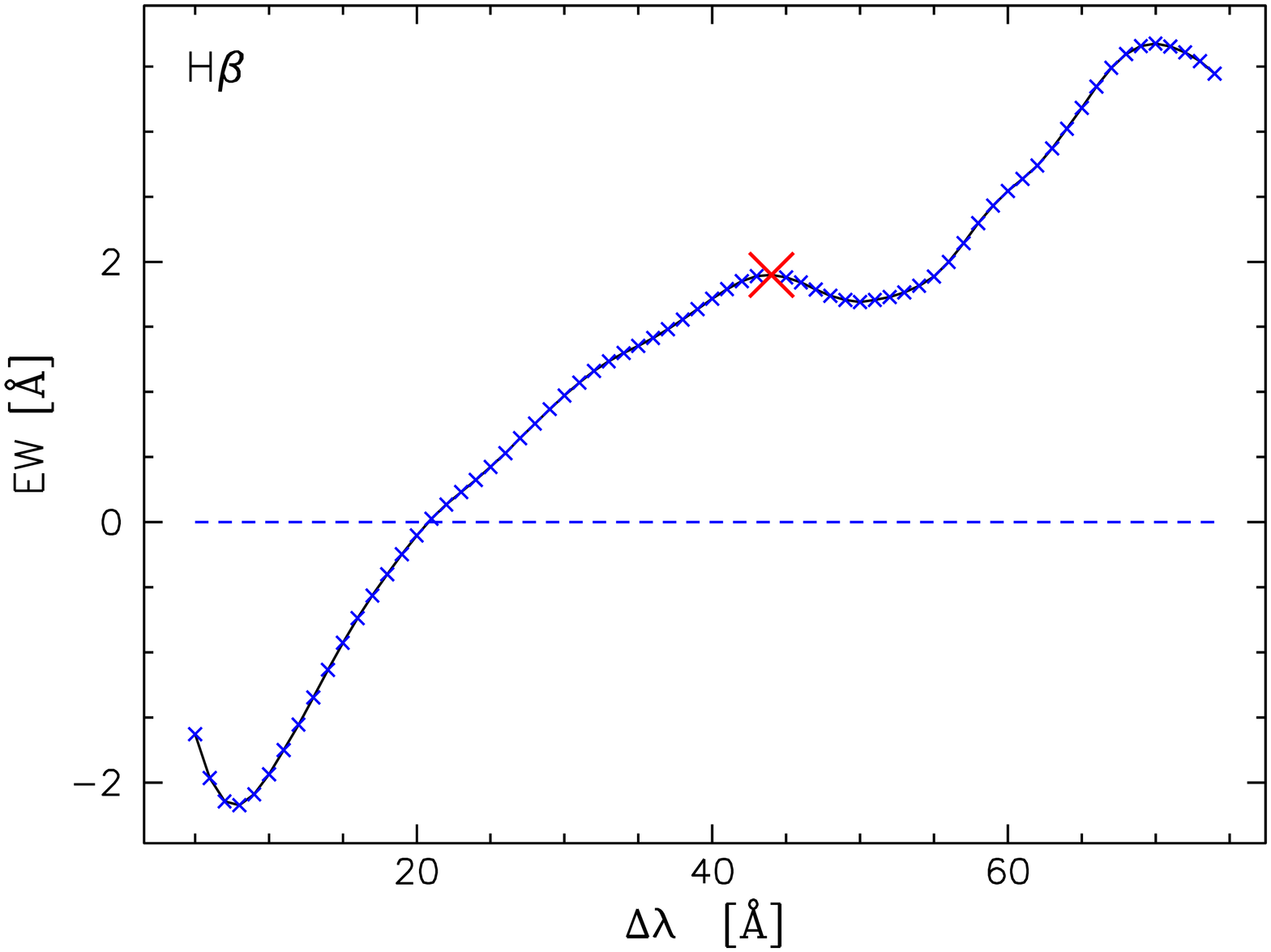}} \\
\end{tabular}
\caption{Here we visually illustrate how our EW measurement method works
on an observed \Hb \ line: in the {\it upper panel} we show how the EW 
is iteratively measured at increasingly larger $\Delta \lambda$
(red, straight line) which are also used to define the continuum
level. The dashed regions correspond to the area measured for the EW 
determination. The analysis of the EW trend curve, plotted on the {\it lower
panel}, provides the correct observed value, shown by the red cross.  
In this case, in which both emission and absorption
components are present, it corresponds to the second zero of the first
derivative of the trend curve itself.}
\label{fig:ewmeas}
\end{figure*}

To find the proper EW value, the line is first measured over a very short 
interval $\Delta \lambda$, under--sampling the real value; then, 
various values of the
EW are computed as a function of $\Delta \lambda$, which is increased
at steps of 1 \AA, starting from an initial width of 4 to 8 \AA,
depending on which line is measured, up to $\sim 80$ \AA. The upper
panel of Fig. \ref{fig:ewmeas} illustrates this process. Here the red straight
lines represent the continuum level defined over increasingly larger 
values of $\Delta \lambda$, and the dashed regions represent the area
where the spectral line is measured (and whose profile is highlighted
with a blue line).

The correct value of the EW is found, among those measured in this
way, by analysing the EW trend curve (see Fig. \ref{fig:ewmeas}, 
lower panel). In fact, the absolute value of the EW
will, in general, monotonically increase as the $\Delta \lambda$
increases, since the line profile is increasingly better
sampled. The point where the trend curve has its first change in slope (first
derivative equal to zero), is generally where the line should ideally 
be measured, both for emission and absorption lines. In spectra
with high Signal--to--Noise (S/N, hereafter), or in theoretical SSPs
spectra, an asymptotic behaviour of the trend curve is also providing
the correct EW value.  An important exception is the case of an
absorption + emission profile, which can be easily found in lines such
as \Hb, \Hg \ or in some cases even in the higher order lines. In this 
case, the first change in the slope of
the trend curve will happen in correspondence to the change from
sampling the emission and the absorption profile. When this is the
case, the second change in slope identifies the correct EW value (see
Fig. \ref{fig:ewmeas}, were such an example is given).

Hence, in order to recognise the absorption+emission pattern, it is first
checked whether the EW trend curve starts with negative or positive
values, for those lines that can be both measured in absorption and
emission (hydrogen lines). If small values of $\Delta \lambda$ yield
negative EW, then we are very likely in the presence of an
emission+absorption pattern, and the EW determination is treated
accordingly.

The \Ha \ line is treated in a different way: due to the close
proximity of the two Nitrogen forbidden lines  ($\lambda=6548$ and
$6584$ \AA, respectively), when the line is in emission the trend
curve analysis may become too complicated and impossible to treat with this method. 
This is particularly true
in a number of different, various situations such as when the lines
are very bright, or if the gas emitting the lines has multiple
velocities components, so that the three lines' profile can be
partially blended, but also for medium resolution spectra, like
the ones we are dealing with. To properly deal with such issue, we
first measure the line within a small ($\sim 15$ \AA) $\Delta\lambda $
interval; in this way we check whether the line is in absorption or in
emission. In the first case, the trend curve analysis method is
applied, as for the other lines. In the second case, a fixed $\Delta
\lambda\sim 70$ \AA \ is adopted, so that this measure will safely
include both \Ha \ and the two Nitrogen lines.

We stress here that this method has been tweaked in order 
to deal with spectra dominated by an emission component of stellar origin. 
Of course, we can also properly measure lines in LINERS or type-2 AGN,
at least for the few ones found in \wings \ (an adjustment to the maximum
value of the EW might be needed to deal with the high--luminosity ones). 
The measurements of (emission) lines in type-1 AGN is instead a much
more complicated issue, as in that case permitted lines can have FWHM of
several thousands of km/s, and they easily end up blending with 
neighbouring ones (e.g. \Hb+\Oiii). Our method is currently not 
capable of properly measuring lines in such spectra, which are 
anyway extremely rare in our sample, hence not affecting by 
any meaning the scientific analysis.

In Table~\ref{tab:lines} we summarize the rules used for the choice
of the most appropriate value of $\Delta\lambda$. In general, when only one 
of the profiles (i.e. emission or absorption) is present, the first change
in the slope of the EW trend curve represents the best guess for the
continuum determination. If both the emission and absorption components
are instead present, the $\Delta\lambda$ corresponding to the second
slope change is instead chosen.

We have chosen a sample of spectra of different types, spanning
from early to late-type, and with different signal-to-noise, and we 
have measured, using the package \texttt{splot} within {\sc iraf}, 
the equivalent widths of the most significant lines. The comparison
between the automatic and the manual methods are shown, for a
subsample of 4 lines, in Fig.3 of \cite{fritz07}, but the results are 
very similar for the other lines. In summary, we found that our 
automatic method returns EW values
which are in agreement, within the estimated errorbars, with those
that have been manually measured. Furthermore, the detection of 
emission lines is reliable for observed spectra with a signal--to--noise
values down to $\sim 5$. Some subjectivity can, of course, affect the
choice of the underlying continuum especially in spectra with the lowest
quality, but this is then reflected in a larger errorbar.
 
\subsection{Uncertainties}\label{sec:ewerror}
To give an estimate of the errorbars associated with an EW
measurement is not straightforward. We consider the main source of
uncertainty to be the choice of the continuum. As already pointed
out, the EW measurement is quite dependent on the fluctuations of the
nearby continuum, hence, on the local S/N. To account for this, 
we measure the equivalent
width of the line at a $\Delta \lambda$ that is, respectively, 5 \AA \
(this number being defined empirically, by checking the lines'
measurements on observed spectra at different S/N) larger and smaller
than the value found by the automatic tool, and take the
semi-difference of these two values as the errorbar. Note that, by
adopting such a definition for the uncertainties on the definition of
the spectral continuum, the better the latter is defined, the smaller
the measured uncertainty will be. Even though this is the dominant
source of uncertainty, we also account for a ``poissonian'' error,
given by:
\begin{equation} 
\Delta(EW)=0.5\times \sqrt{|EW|}
\end{equation}
which is added in quadrature to the measurement
uncertainty. Errorbars computed in this way are in very good agreement
with those formally computed by propagating the measured errors.

As a further check, we have used the subsample of \sdss \ galaxies with 
spectroscopic data which are in common with our survey \citep[see also
the comparison performed in][]{fritz11} to compare the EW values and
see whether they agree within the uncertainties calculated by the 
measurement method. Of the 395 objects with spectroscopic data 
from both surveys we use, for this comparison, the $\sim 250$ which 
are brighter than M$_V=-18$ (see Sect.\ref{sec:NS}). Before proceeding, 
an important caveat must 
be considered: there is a difference in the size of the fibers used in 
taking \wings \ and \sdss' data, and this means that the physical regions we
are observing might include, in some cases, slightly different stellar 
populations. Added to this, the location of the fiber might also not be exactly 
the same and the signal--to--noise of some of the spectra can also be 
very different. All these effects might increase the scatter in the comparison.

Despite this, we find that the EW values measured for the two different
datasets, and in particular those of \Oii \ and \Hd \ (which are those upon
which our classification scheme is based), very well agree within the given
uncertainties. The rms of the relation comparing \Oii\ measurements is $\sim 6$ \AA,
which is in fair agreement with the average errorbar of $\sim4$ \AA, especially if 
we take into account the aforementioned caveats on the fibers size and 
position. In the case of \Hd, the rms is $\sim 0.8$ which again is 
consistent with the average errorbar of $\sim 0.7$ \AA. This not only 
makes us more confident that the EW 
uncertainties are well estimated, but it further strengthen the reliability 
of the spectral classification.

\subsection{Noisy detections and caveats}\label{sec:noisy}
One of the trickiest issues concerns the detection of lines in
spectral regions with low S/N or in the proximity of the telluric
absorption bands which are sometimes not properly subtracted from the
spectra. First of all, we do not want the measurements, i.e. the
analysis of the EW trend curve, to be influenced by the fluctuations
produced by noise and, second, we require that noise fluctuations and
spikes are not interpreted as real spectral lines. The blue range of a
spectrum contains, for typical \wings \ spectra and redshifts, 7
relevant lines within a 400 \AA \ range and, unfortunately, is also
the most prone to flux calibration uncertainties and the most affected
by noise. To overcome this issue, some empirical criteria are defined
to recognise noisy patterns and to distinguish them from real spectral
lines:

\begin{itemize}
\item[$\bullet$] the EW of forbidden lines can only be negative
(i.e. in emission); positive values are hence interpreted as the line
being absent or the spectrum highly noisy in that region;
\item[$\bullet$] late Balmer lines, such as \Hd \ or H$\epsilon$,
can be measured in emission if and only if other lines --such as for
example, \Oii, \Oiii, \Hb \ or \Ha \ which are typically brighter-- 
are also detected in emission;
\item[$\bullet$] an upper limit is set to the values of both
absorption and emission lines: EWs of absorption lines cannot be
higher than 16 \AA, a value derived from SSP models. When emission lines
are taken into account, the maximum (absolute) value that a line can
assume is dependent on the line itself, and was computed from SSP
spectra. Lines having EW outside these ranges are considered as noise.
\end{itemize}

The shape and, more in general, the characteristics of the EW trend curve 
can depend on the spectral resolution. In fact, especially in low 
signal--to--noise and high resolution spectra, the trend curve can assume an
erratic behaviour, particularly at a local level. Thus, the method had to be
slightly tweaked so to take into account for a possible sensitivity to the 
spectral resolution. This was done by testing and calibrating the algorithm on 
spectra having the same resolution of those in the \wings \ datasets. Furthermore,
as described in \S\ref{sec:ewerror}, we have applied our method to a set of \sdss \
spectra, and found that it does not show a strong sensitivity on the 
resolution, at least as far as datasets with such characteristics are considered. 

Finally, we consider as unreliable values above -2 \AA \ for the EW of \Oii \
and \Oiii. These cases are identified with a 0.0 in the catalogs.

\section{\wings \ spectra}\label{sec:spec}
The \wings \ spectroscopic sample was built up from two sets of
observations: spectra of galaxies in clusters at a negative
declinations were taken with the 2dF multifiber spectrograph at the
Anglo-Australian Telescope, and constitute the ``south sample''.
Spectra in the ``north sample''  were instead taken
at the William Herschel Telescope using the AF2/WYFFOS (see \S
\ref{sec:data} for more detailed information about the spectra).

Due to both technical issues and bad weather conditions, a number of
spectra taken at the William Herschel Telescope turned out to be
of very poor quality, and had to be discarded as not suitable for our 
work. In this way, the number of spectra of members for some 
clusters had been so drastically reduced, that we decided not to include 
these clusters in our analysis. In Appendix~\ref{sec:quality} we provide 
more details both on the issues and on the adopted rejection/acceptance 
criteria.

\begin{table}
\centering
\begin{tabular}{lrrr}
\hline
\sc{cluster}&\sc{total}  & \sc{members} &  \sc{redshift} \\
\hline
    A1069   &        97    &       39   &    0.0653 \\
      A119   &      241     &     156   &    0.0444 \\
      A151   &     235    &        85   &    0.0532 \\
  A1631a   &     184    &       109   &    0.0461 \\
    A1644   &     225   &       165   &    0.0467 \\
    A2382   &     209  &       134   &    0.0641 \\
    A2399   &     191  &       110   &    0.0578 \\
    A2415   &     180  &        95   &    0.0575 \\
    A3128   &     274  &       196   &    0.0600 \\
    A3158   &      263   &      172   &    0.0593 \\
    A3266   &      247   &      215   &    0.0593 \\
    A3376   &      130   &       86   &    0.0461 \\
    A3395   &      178   &      121   &    0.0500 \\
    A3490   &       186   &      69   &    0.0688 \\
    A3497   &      135   &       74   &    0.0680 \\
    A3556   &      155   &      105   &    0.0479 \\
    A3560   &      156  &       107   &    0.0489 \\
    A3809   &       165  &       95   &    0.0627 \\
     A500    &      133    &      89   &    0.0678 \\
     A754    &     141   &       119   &    0.0547 \\
    A957x   &      116   &       62   &    0.0451 \\
     A970    &      172    &     114   &    0.0591 \\
    A1795   &        57   &      29   &    0.0633 \\
    A1983   &        66   &      31   &    0.0447 \\
    A2457   &       50    &      35   &    0.0584 \\
    A2626   &       41   &       27   &    0.0548 \\
     A376    &      58    &       43   &    0.0476 \\
    Z8338   &        60   &      37   &    0.0494 \\
    Z8852   &       36   &       25   &    0.0408 \\
\hline\T
\sc{total}   &    4381  &     2744 &                    \\
\hline
\hline
\end{tabular}
\caption{List of the clusters used for this work. For each cluster
we report the total number of spectra for which 
we have reliable EW measurements, those which are confirmed 
cluster members, and the cluster's redshift.}
\label{tab:cl_summ}
\end{table}
This selection reduces the number of spectra that are used in this
work to 4381 out of the $\sim 6000$ that were originally observed, 
of which 2744 are spectroscopically confirmed members,
belonging to 29 clusters in both the south and north sample. In 
Table~\ref{tab:cl_summ} we report the cluster names, the total number of
spectra and the confirmed members, and their redshifts.

\subsection{Spectral classification}\label{subsec:spclass}
Adopting the spectral classification defined by
\cite{dressler99} (MORPHS collaboration), which was developed to study
distant galaxy clusters, we divide the spectra of our sample into 6 classes,
based on the EW of \Oii \ and \Hd, according to the scheme
presented in Table \ref{tab:ewclass}. Spectra for which none of the two
lines was detected are classified as {\it noisy}. The broad physical
meaning of such a classification is discussed in detail by
\cite{poggianti99} and \cite{poggianti09b}; 
we summarize it briefly in the following:
\begin{itemize}
\item {\it e(a)} emission-line spectra with strong \Hd \ in absorption, 
signature of the presence of A--type stars. They are typical either
of dusty starbursts, or of systems where an episode of substantial
star formation was abruptly interrupted, and only a residual activity
might be currently present;
\item {\it e(b)} emission--line spectra with very strong emission lines, 
typical of star-bursting systems with low to moderate dust extinction;
\item {\it e(c)} emission--line spectra with moderate-to-weak emission and
moderate-to-weak \Hd \ in absorption, typical of a regular (non-starbursting) 
star formation pattern, as those distinctive of ``normal'' spiral galaxies;
\item {\it k} spectra resembling those of K-type stars, lacking emission lines
and with weak \Hd \ in absorption,
typical of passively evolving galaxies with neither current 
nor recent star formation activity. Such spectra are normally found in elliptical
galaxies;
\item {\it k+a/a+k} spectra displaying a combination of signatures typical of both 
K and A --type stars, with strong \Hd \ in absorption and no emission lines,
typical of post--starburst/post--starforming galaxies whose star formation
was suddenly truncated at some point during the last $0.5-1$ Gyr.
\end{itemize}

Following this broad classification, all emission lines galaxies are
those which are undergoing a process of star formation at the very
moment of their observation. Another origin of emission lines is also
possible and it will be discussed later on. The other types represent  ``passive''
spectra, the {\it k} being dominated by old stars (older than $\sim 2$
Gyr), and the {\it k+a} and {\it a+k} showing evidence for an
increasingly important presence of A-class stars, which are  commonly
recognised as a signature of a relatively recent ($<1$ Gyr) burst or at least
activity of star formation.

\begin{table*}[!t] 
\centering
\begin{tabular}{lccccccccr}
\sc{sp.ty.} & cat \sc{id} &   \Oii             &      &   \Hb       &   &   \Oiii   &  &   \Hd  & N (\%) \\
\hline
\hline
e(a)          &         1         &       $<0$      & {\sc or}     &     $<0$            &  {\sc or} & $<0$  &  {\sc and}  & $\geq 4$      &    6.9 \%    \\
e(b)          &         2         &      $<-40$    &{\sc or}     &     $<-12.5$     &              &    ...     &{\sc and}    &    $<4$          &    2.5 \%    \\
e(c)          &         3         & $-40$ to 0  &{\sc or}     & $-12.5$ to 0&{\sc or}  & $<0$   &{\sc and}    &    $<4$          &  29.9 \%    \\
k               &         4         &       $>0$       &       ...       &          ...           &       ...       &       ...       &{\sc and}    &    $<3$          &  49.6 \%    \\
k+a          &          5         &       $>0$       &       ...      &           ...           &       ...       &       ...       &{\sc and}   &    3 to 8   &  10.4 \%    \\
a+k          &          6         &      $>0$        &       ...      &           ...          &       ...       &       ...       &{\sc and}   &    $\geq8$    &     0.7 \%     \\
\hline
\end{tabular}
 \caption{Summary of the criteria adopted to classify the
 spectra based on spectral lines criteria. Other lines (namely \Hb \
 and \Oiii) were introduced with respect to previous works, in order
 to be able to classify a spectra in those cases when the \Oii \ line
 was out of the observed spectral range or overwhelmed by noise. 
The last column presents the spectral type fractions, 
corrected for incompleteness, among galaxies to which a spectral 
classification could be assigned.}
\label{tab:ewclass}
\end{table*}

Note that in few cases, in order to distinguish between the 3 
emission line classes, we used, if possible, also \Hb \ and \Oiii. 
These two lines were used in those cases, mostly for spectra of
the north sample, for which the \Oii \ line either was out of the
observed range, or it was non-detected (the blue range of the spectra
is often quite noisy). The use of \Hb \ in particular, has been
already exploited by \cite{yan06} as a mean to distinguish and
properly classify \ea \ and {\it k+a} spectra. We searched for a
relation between either the EW of \Oiii \ (5007 \AA), which is easily
detected --when present--, or that of \Hb, and the EW of the
\Oii. This relation is intended to be used only to distinguish between
the \eb \ (\Oii$<-40$ \AA) and the \ec \ (\Oii$\geq -40$ \AA) classes,
while the classification as an \ea \ is made mainly by means of \Hd. 

No clear trend is found for either of the two lines and, if we average
the EW value of \Oiii \ and \Hb, for those galaxies where
EW(\Oii)$\leq -40$, we obtain $-12.5$ and $-32.5$ \AA, for \Hb \ and
\Oiii, respectively. We choose to use \Hb, as a proxy for the
\Oii \ line, since it has the lower dispersion (even though we are
conscious that this is in part due to the fact that also the average
value of \Hb \ EW is lower with respect to that of \Oiii). Table
\ref{tab:ewclass} is meant to summarize the criteria used to assign
the spectral classification.

Note that the spectrographs at both telescopes allow us to observe, 
for all \wings \ clusters, the spectral region containing the \Oii \ line, 
which is crucial for the classification scheme we have chosen. In the worse
case, at the cluster's average redshift, the \Oii \ line is observed at 
$\sim 3880$ \AA, well above the lower limit of 3600 \AA. On the other side,
most of the spectra in the north sample do not encompass the \Ha \ line which
is anyway not used for spectral classification purposes, hence not affecting any of 
the derived quantity we use for the present work.

\subsection{Compatibility between south and north sample classification}\label{sec:NS}
\begin{figure}
\centering
\rotatebox{270}{\includegraphics[height=.5\textwidth]{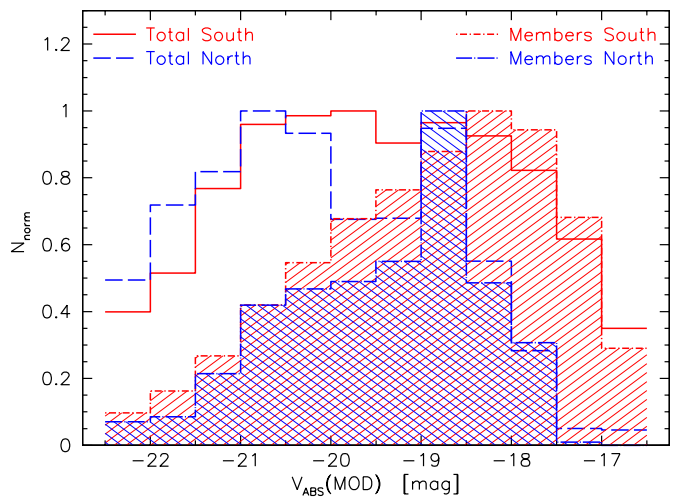}} 
\caption{Distributions of absolute V magnitude for galaxies in the north 
and south samples (different colors and lines). 
Absolute magnitudes are computed as the
integral convolution between the rest-frame model spectrum and the V
band transmission curve. All distributions are normalised to the their peak
number, and both magnitude and geometrical completenesses are taken
into account.}
\label{fig:magstat}
\end{figure}

The V-band absolute magnitude distributions are shown in
Fig. \ref{fig:magstat}, for the north and south samples separately, 
distinguishing
also the spectroscopically confirmed members. Absolute magnitudes used
for this work are those computed based on the best-fit model to the
observed spectrum \citep[see][]{fritz11}, so that the {\it
k-correction} is automatically taken into account, and no assumptions
need to be made. 
We note that at the faint end the north distribution declines
at magnitudes brighter than the south sample (M$_V \simeq -18.5$), 
probably due to severe incompleteness setting in at brighter magnitudes
in the north. Nevertheless, as the north subsample
contains less than 10\% of the total \wings \ members, 
this will not alter the significance of our results. 
For these reasons, we can carry on our analysis of the whole
sample using galaxies with M$_V < -18$. 

\begin{figure} 
\centering
\includegraphics[height=.51\textwidth]{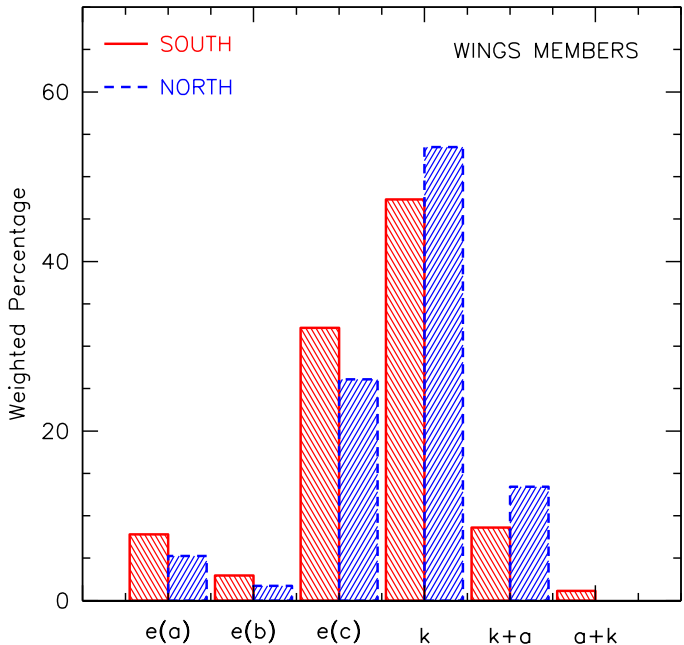}
\caption{Spectral types distribution of \wings \ spectroscopically
confirmed members. South and north
samples are shown separately with different colours and lines, for
comparison. A cut at absolute magnitude M$_V=-18$ has been applied.}
\label{fig:ewclass}
\end{figure}

In Fig. \ref{fig:ewclass} we show the distribution of the various
classes within the north and south subsamples for member
galaxies. Following what was previously discussed, a cut in 
absolute magnitude at M$_V=-18$ was applied. We
choose to show only cluster members, since the
non-members population contains a mix of field galaxies and foreground
structures, which we are not able to properly distinguish. 
The percentage of the
various classes are computed taking into account both magnitude
and geometric completeness. 

As described in Appendix~\ref{sec:quality}, the EW measurements had 
to be performed manually for some spectra, and this could, at least in 
principle, make the measures --and hence the classification-- non 
homogeneous between the south and north subsample, possibly 
introducing some bias. 

From Fig. \ref{fig:ewclass} we note that the relative
number of the various spectral types is similar in the two samples,
meaning that, once completeness corrections are taken into account,
and as long as galaxies brighter than $M_V=-18$ are considered, the
two samples can be considered as homogeneous and comparable, 
within the statistical uncertainties.

\section{The characteristics of the local clusters galaxy population}\label{sec:properties}
We now analyse the properties of the galaxies in our sample, with a
focus on their spectral classification and related characteristics.

\subsection{General properties}\label{sec:general}
We were able to successfully classify about 90\% of the spectra of our
survey (both cluster members and background objects), based on \Oii \ and \Hd
\ (or \Hb, when the \Oii \ lines was not observed, see Sect. \ref{subsec:spclass}). 
Spectra without a classification either have 
a very low S/N, or the lines were not measured due to strong artifacts
affecting their profiles. 

With this information, we can now analyse the properties of the cluster 
population, discussing the characteristics of the various spectral types 
as a function of their physical properties such as luminosity, morphology, 
stellar mass, and position within the cluster. To properly do this, we limit our sample so that
it includes only galaxies with a total absolute magnitude equal or
brighter than $-18$ in the V band (see Sect.~\ref{sec:NS}). With this cut 
in magnitude, we are left with 2296 out 
of the original 2744 in the members' sample. From now on we will consider 
only objects in this magnitude limited sample, unless otherwise stated. 
We stress here that at this magnitude limit the south sample is complete at
the $\sim 70$\% level \citep[see the discussion in][]{cava09}.

Table~\ref{tab:maglim} summarizes the number distribution of the three main 
morphological types. Galaxy morphologies are taken from \cite{fasano12}
and have been obtained with MORPHOT, an automated tool designed to simulate
as closely as possible a visual morphological classification. MORPHOT uses 
a combination of 21 imaging parameters complementing the classical 
Concentration/Asymmetry/clumpinesS indicators, derived from the V-band
\wings \ images, and provides a parametrical morphological classification 
based both on a maximum likelihood technique 
and on a neural network machine \citep[see][for details]{fasano12}.
This piece of information will be used later on (see, e.g., Sect.~\ref{sec:morph}). 

\begin{table}
\centering
\begin{tabular}{cccc}
{\sc total} & {\sc ellipticals}  & {\sc S0}  & {\sc spirals}  \\
\hline
2296        & 653                     &     1052     &      574           \\
    ...          & 28.4\%               &     45.8\%  &     25.0\%      \\
\hline
\hline
\end{tabular}
\caption{Summary of all the morphological types, identified as cluster members, 
in our magnitude limited sample (corrections for incompleteness are here not
taken into account). BCGs are not included.}
\label{tab:maglim}
\end{table}
Considering the spectral classification as outlined in
Sec. \ref{subsec:spclass}, we found that the galaxy population in
local clusters is dominated by the \k \ spectral type, including about
50\% of the galaxies in our spectral sample, followed by the \ec \
class ($\sim 30$\%). The post--starburst classes (\ka \ and \ak) 
represent about 11\% of all the galaxies while the two other 
emission-lines classes, \ea \ and \eb, contain 7 and 3\% of all members, 
respectively (see Table~\ref{tab:ewclass}). These fractions have
been calculated taking into account completeness corrections.

\subsection{The V-band luminosity distribution}\label{sec:lumfunc}
\begin{figure*}
\centering
\begin{tabular}{ll}
\includegraphics[height=0.65\textwidth]{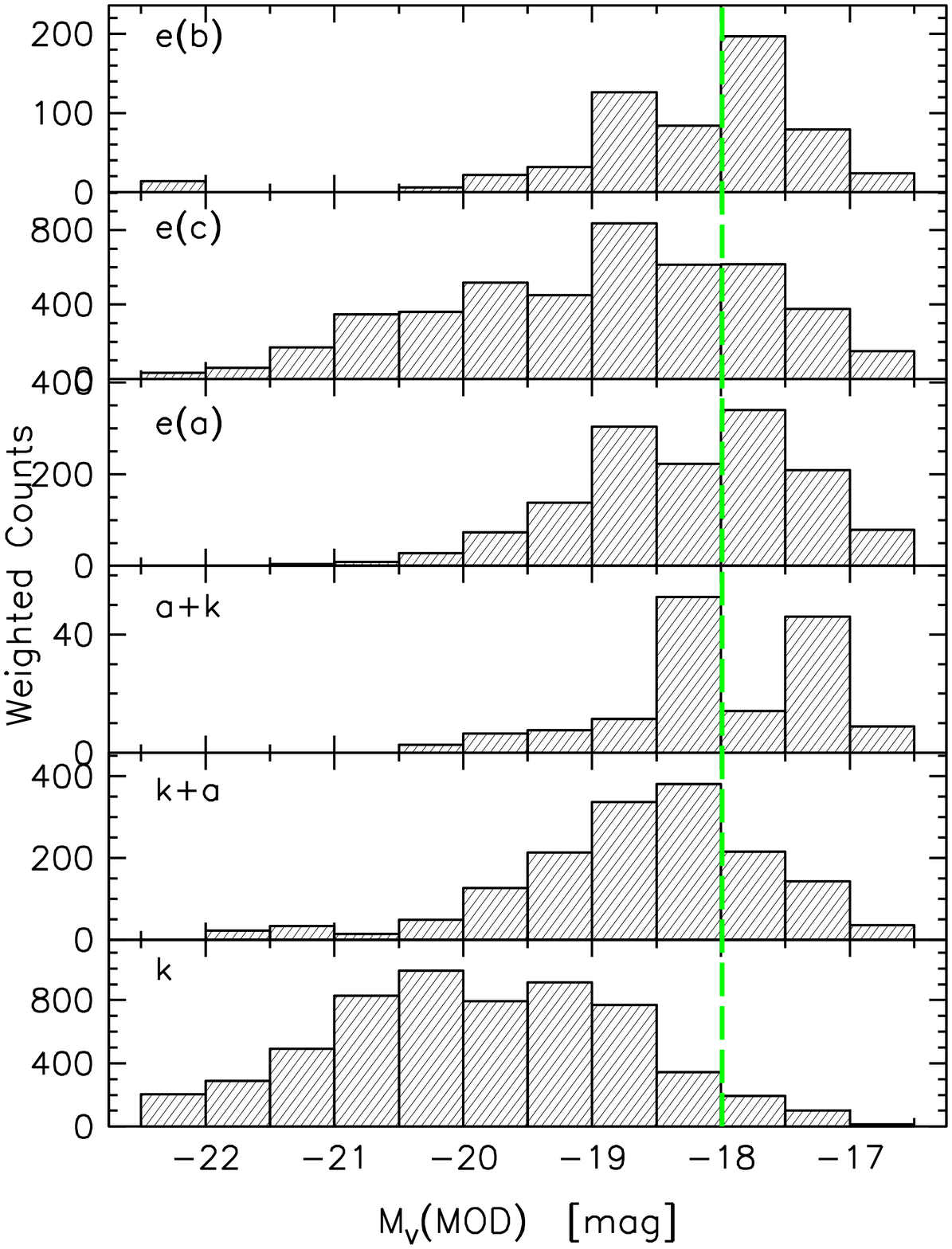} &
\includegraphics[height=0.65\textwidth]{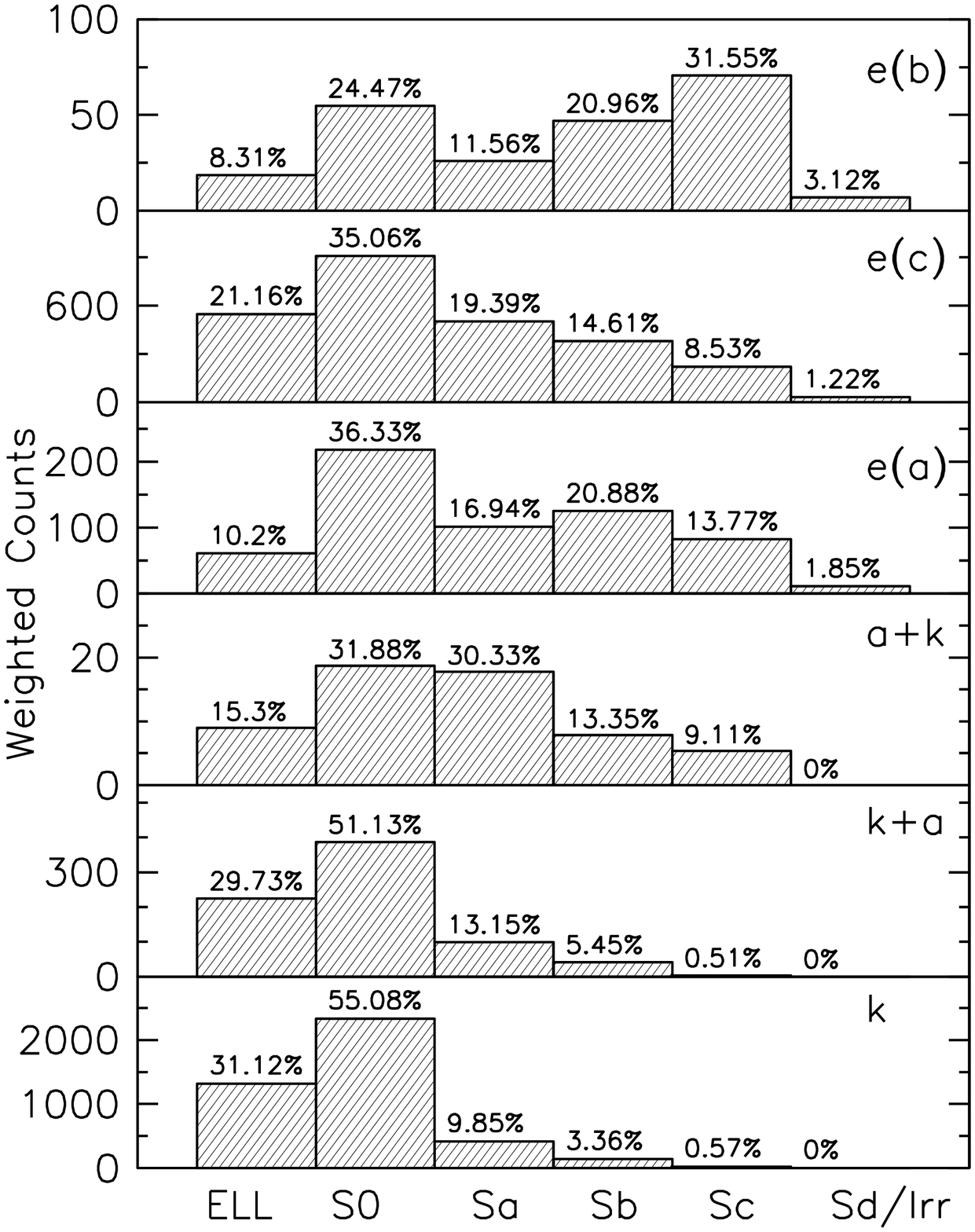} \\
\end{tabular}
\caption{{\it Left panel}: shaded histograms represent the number
of galaxies in the 6 spectral types, as a function of the
absolute V magnitude; the green--dashed line marks the magnitude
limit of the subsample used in the analysis (for this plot only, we consider
no cut in magnitude). Galaxies brighter than 
M$_V=-22.5$ are included in the brightest bin. {\it Right panel}:
occurrence of morphological types as a function of the spectral class. The
percentages refer to the number of objects with a given morphology
with respect to each spectroscopic class. For both plots data have
been weighted by both magnitude and geometric completeness. For the
right--hand plot only galaxies with M$_V < -18$ have been considered.}
\label{fig:sptype}
\end{figure*}

On the left panel of Fig. \ref{fig:sptype}, we compare the
distributions of the V-band absolute magnitude for the various 
spectral classes. Note that this magnitude was computed from
the best fit model to the observed spectrum \citep[see][]{fritz07,fritz11}.

The V-band luminosity distributions span a range of 3 to 5 magnitudes, 
depending on the spectral type, peaking in general at different magnitudes and
displaying also a different median luminosity as a function of the spectral class.

It shows no significant differences between the \ea \ and the \eb \ spectral 
classes, peaking around $M_V\sim-18$ for both types, having the same 
median luminosity and very similar range. 
The luminosity function for the other emission-line type, the \ec \ galaxies, 
substantially differs in that not only it displays a tail towards brighter 
magnitudes, but it also peaks at higher luminosities ($M_V\sim-19$; note 
this is above the magnitude limit we consider).

As for the distributions of the two post--starburst classes \ka \ and \ak, 
they both peak around $M_V\sim-18$, even though, especially in the 
case of the \ak \ class we are most likely limited by the poor statistics. 
They differ, though, with respect to the 
range in luminosity which is broader for the \ka \ class, reaching 
$M_V\sim-22$, almost matching the high--luminosity tail seen for the \ec \ and
\k \ galaxies. The characteristics of the luminosity distribution we find for \ka's , 
are very similar to those of the ones in the local, massive Coma cluster 
\citep[see][]{poggianti04}, lacking a significant bright population, 
and having a similar typical absolute magnitude of $\sim -18$.

The V-band luminosity distribution of
galaxies with a \k -like spectrum is yet clearly different with respect to all the
other classes in basically all of its characteristics: peaking at
$M_V \sim -20.5$, they also have the most luminous tail in the
distribution while at the same time lacking, when compared to the other classes, 
a significant population
of objects with absolute magnitudes lower than $M_V \sim -19$. In fact, together
with the \ec \ type, this is the only class whose luminosity function decreases 
before our magnitude limit cut.
\begin{figure}
\centering
\includegraphics[height=0.65\textwidth]{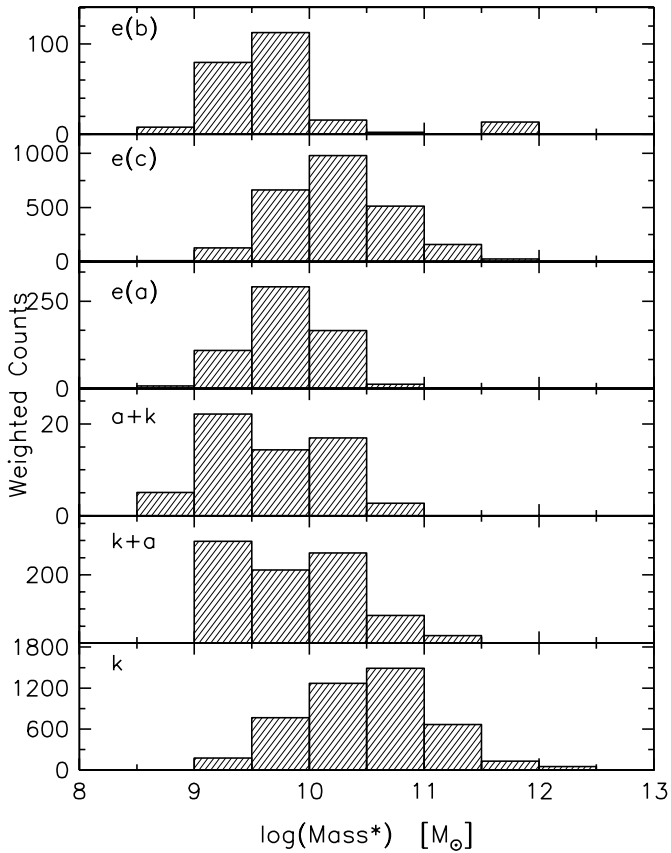}
\caption{Stellar mass distribution for the 6 spectral types, in the
magnitude limited sample, and for objects within R$_{500}$.}
\label{fig:mass_sp}
\end{figure}

\subsection{Morphology distribution of the spectral types}\label{sec:morph}
An analysis of the morphological mix of the \wings \ galaxy population has
already been carried out in \cite{poggianti09}. Here we focus on the 
relation between morphology and spectral type.
 
In the right-hand panel of Fig.\ref{fig:sptype} we show how the spectral 
classification correlates with the morphology for our magnitude--limited sample. 
\wings \ galaxies follow in general the broad relation between 
morphology and star formation history.

Globally, the dominant morphological type is represented by S0s, 
accounting for more than 50\% of both the passive spectral type \k \ 
and the post--starburst \ka. They are also the most numerous 
among every spectral class with the exception of the \eb,
which is instead dominated by late-type spirals (Sc). More than 80\%
of the passive galaxies \k \ are early-types (E+S0), and the same 
goes for the the post--star forming \ka \ class. 

On the other side, the majority of \ea, \eb \ and \ak \ are spirals. While the
morphological fractions of \ea \ and \ak \ are similar, the \eb \
class clearly stands out being dominated by spiral galaxies, with more 
than the 50\% of objects being found among later types than Sb.
The \ec \ spectral type is almost equally populated by early 
(E+S0) and late types (Sp). 

The presence of the various morphological types among the \k 
\ and \ka \ galaxies is very similar: we observe a clear prevalence 
of ellipticals and S0s, no Sd/Irregular, and a fraction of spirals 
less than 20\% in both classes. Furthermore, this fraction is 
monotonically decreasing when going from \ak, to \ka \ and finally to 
\k, where the relative occurrence of Spiral galaxies is the lowest, 
probably mirroring the transition from a post--star forming
phase to a passive evolution pattern.

Finally, there seems to be a significant difference
in the morphology distribution between the two so--called post--starburst 
classes, \ka \ and \ak, with the latter distribution resembling more that 
of the spiral types (even though it should be pointed out that the statistic is
not optimal for the \ak \ class).
A similar result was found by \cite{dressler99} at higher redshift.

To summarize, while there is indeed a broad correspondence between 
the morphological and spectral classifications, we find that galaxies in 
local clusters lack an {\it univocal} equivalence between their morphology 
and their spectral characteristics, a result 
already well known at high redshift \citep[e.g.][]{dressler99,couch01}, 
meaning that the evolution of these two properties is, at least to some 
degree, decoupled.

\subsection{Emission line galaxies in local clusters}\label{sec:emlines}
As already described in Sect.~\ref{sec:lumfunc}, the V-band luminosity
distribution is very similar for the \ea \ and \eb \ spectral type, 
but substantially different for the \ec. This somehow reflects 
on the stellar mass distribution \citep[see Fig.~\ref{fig:mass_sp}; mass 
values are taken from the catalog presented in][]{fritz11}: \ec \
galaxies peak at around $10^{10.5}$ M$_\odot$, while the distribution
for \ea \ and \eb \ peaks at stellar mass values 0.5 dex lower, with the \ea \ 
population having, in proportion, a slightly higher fraction of massive 
galaxies with respect to the \eb's;
otherwise the distributions of the latter two spectral types are very similar.

It should be said
that, as we noted in \cite{fritz07}, the more a galaxy has been
actively forming stars in a recent past, the more difficult it is to have a robust 
estimate of the stellar mass of the oldest stars from optical spectra alone, 
given their high 
mass--to--light ratios. Thus, in principle, the mass of  \eb \ galaxies 
is the one potentially more prone to be underestimated when calculated
with data similar to ours. Nevertheless, a comparison with an estimate 
of the stellar mass calculated using only B and V-band photometry and adopting
the \cite{belldejong01} recipe, gives fully compatible results, 
\citep[see][]{fritz11}. A similar check exploiting K-band photometry
from \cite{valentinuzzi09}, further validates our results.

We now analyse the fraction of star-forming galaxies as a function of
the cluster properties, such as its velocity dispersion and X-ray
luminosity. As done in several previous works \citep[see
e.g.,][]{dressler99,poggianti06}, we use the \Oii \ (3727 \AA) as a
tracer for the star formation activity, so to be able to consistently
compare our results with other studies, especially those at higher redshift. 
Even though this line is prone to reddening issues and its intensity 
depends on both the gas metallicity and physical conditions, 
its presence/absence can be reliably used to discriminate between 
star-forming/passively evolving galaxies. 

For each cluster we calculate the fractions of galaxies,
weighted for completeness,
displaying the \Oii \ line in emission, i.e. having a rest-frame EW
$<-2$ \AA. As the data coverage, in terms of projected radial distance,
is limited by the characteristics of the spectrographs we have used 
for the survey, and since the clusters span a non negligible range in
redshifts, observing galaxies in clusters at different distances means we 
are sampling areas at different distances from the cluster's centre. 
For this reason, in order not to introduce any bias and sample the
galaxy population homogeneously, only galaxies within R$_{500}$ 
(we use the relation R$_{500}\simeq 0.50\times$ R$_{200}$) are 
considered.
 
\begin{figure*}
\begin{tabular}{rr}
\rotatebox{270}{
\includegraphics[height=.480\textwidth]{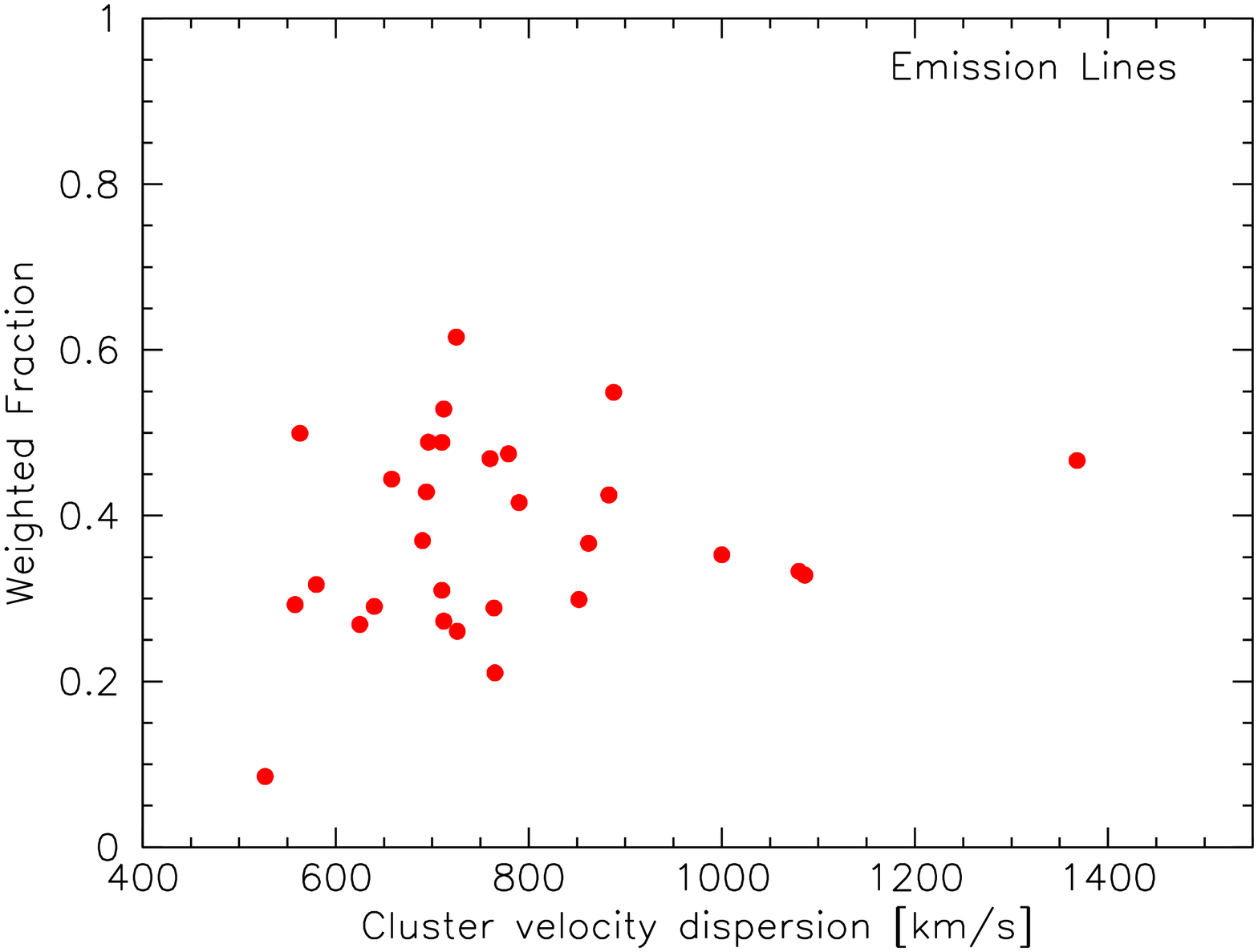}} &
\rotatebox{270}{
\includegraphics[height=.480\textwidth]{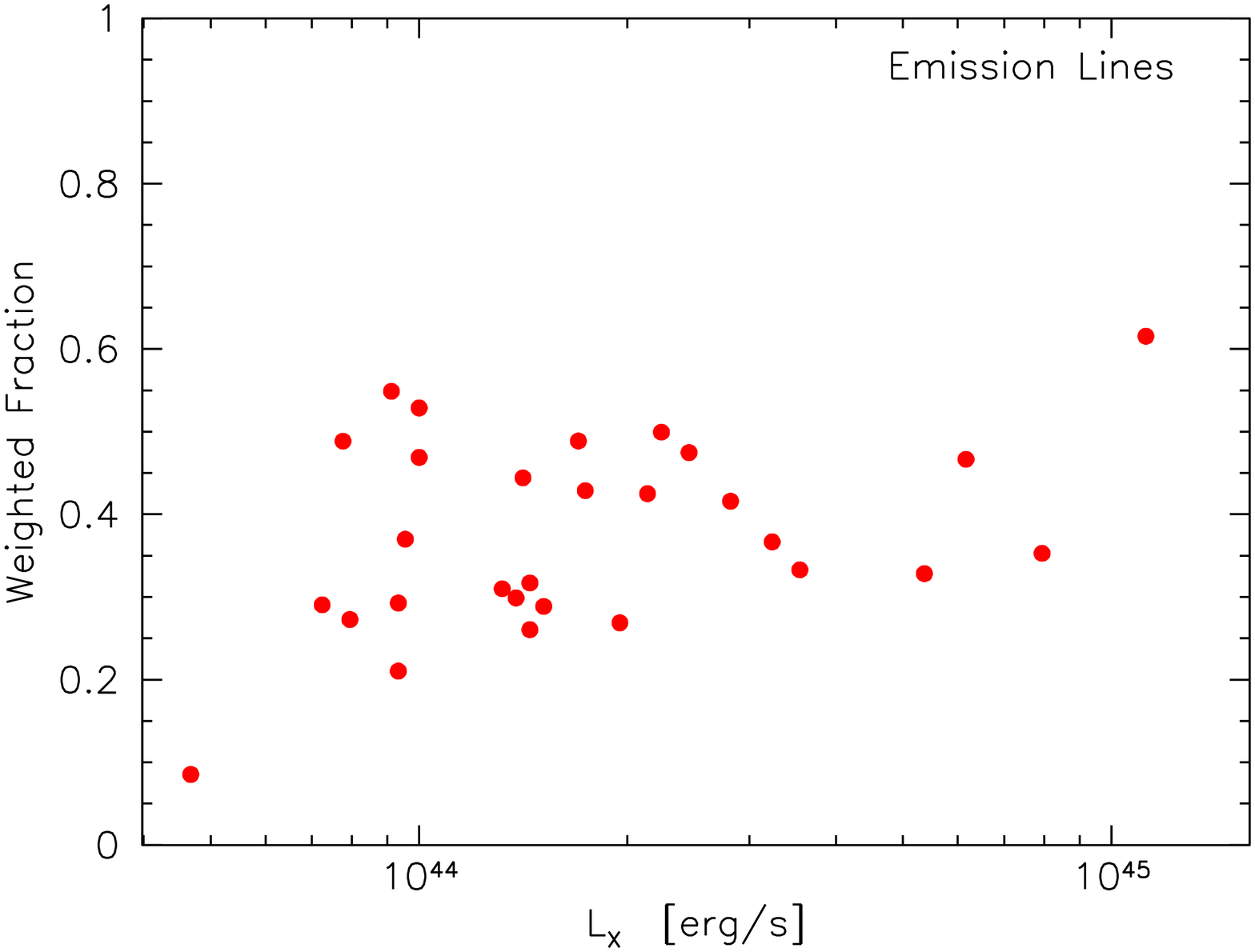}}  \\
\end{tabular}
\caption{Fraction of emission line galaxies (i.e. belonging 
to any of the first three spectral classes defined in Table~\ref{tab:ewclass}) 
as a function of the cluster velocity dispersion ({\it left}) and total X-ray
luminosity ({\it right}). Only galaxies within R$_{500}$ are taken into account 
(see text for details).}
\label{fig:oiifrac}
\end{figure*}

The results are then plotted against the two main global properties 
of the clusters, namely their velocity dispersion and X-ray 
luminosities, and displayed in Fig.~\ref{fig:oiifrac}.

In \cite{poggianti09} we have already analysed the frequency of
morphological types as a function of the cluster velocity dispersion and
X-ray luminosity. Here we perform a similar analysis based on the spectral
classification, where X-ray luminosity (L$_X$) and the cluster 
velocity dispersion ($\sigma$) are used as a proxy of the total cluster mass.

Similar to what found in \cite{poggianti09} we note that there is no, or at
most a weak correlation between the fraction of galaxies with emission
lines and the velocity dispersion, as far as clusters having a velocity 
dispersion in the $\sim 500 \div 1400$ km/s range are
concerned (i.e. the range of \wings \ clusters). This is also in agreement with the 
results of \cite{poggianti06}, who analysed a low-z galaxy cluster sample 
drawn from the \sdss. This result suggests that the global properties 
of a cluster are not the main physical drivers of the build up of
the mixture of spectral types. A similar result is found also for post--starburst 
galaxies (see Sect.~\ref{sec:poststarburst}).

We checked with a Spearman test whether there is any significant correlation
between the emission line fraction and the X-ray luminosity. The test yielded 
a correlation coefficient of 0.26, which indicates that there is none.

The fraction of star-forming galaxies in \wings \ clusters, calculated from
each single cluster, is $\sim 33\pm 10$\%, with a large scatter
at any given $\sigma$ and L$_X$. This fraction is larger than the 23\% 
found in \cite{poggianti06}, but this is likely due to the differences in the 
magnitude limit in these two studies, being deeper in \wings, and the
different limit for \Oii detection (-3 \AA \ for the latter work).

Let us now focus on the characteristics of the \ec \ type galaxy
population in a deeper detail. As we have already stressed in 
Sect.~\ref{sec:general}, these galaxies constitute a major component in
local clusters and their luminosity distribution is quite distinct
with respect to the other emission-lines types: not only it peaks at brighter values
but, most important, it contains a substantial population of brighter
objects, which is instead completely absent from the other two 
emission lines classes. If we look at the properties of the most 
luminous (i.e. $M_V\leq -20$) tail of \ec \ galaxies, and we compare 
them to the less luminous ones we note that:
\begin{itemize}
\item they have fainter \Oii: in most of them the equivalent width of this 
line is $\geq -5$ \AA;
\item they have older ages \citep[both mass- and luminosity-weighted: 
values are taken from][]{fritz11};
\item their spectra have redder (B--V) colour (median value of 1.0 against 0.8);
\item they are on average more massive (a median mass of $\sim 4\times 10^{10}$ 
M$_\odot$, compared to a mean value of $\sim 10^{10}$ M$_\odot$
for the others).
\end{itemize}
In order to check whether this particular subsample is biased by 
noisy measurements of the \Oii \ line which is, for these cases, faint, we
visually inspected the spectra and came to the conclusion that in the
vast majority of cases the detection is real and the measure is reliable.

It is worth stressing that the detection of the \Oii \ line, which is the main
feature used for classifying \ec, distinguishing them from \k,
is not univocally related to the presence of a star forming activity in galaxies,
but can instead have a different origin \citep[see, e.g.][]{yan06}. 
Indeed, when inspected one--by--one, most of the high-mass 
(M$_\star > 10^{10.5}$ M$_\odot$) elliptical and S0 galaxies with an
 \ec \ classification, display
spectral features typical of LINERS or low-activity AGNs, such as
a enhanced {\sc [Nii]} (6584 \AA) emission compared to \Ha. This goes 
along the line of the findings by \cite{yan06}, who pointed out that ``\Oii-only'' 
galaxies, or even those with a high \Oii/\Ha \ ratio, are more common among 
the reddest in their \sdss \ emission-lines sample, so probably the most 
massive ones. They classified as ``post--starburst'', galaxies
displaying a high \Oii/\Ha \ ratio (or even with a \Oii \ detection, but no \Ha, the 
so-called \Oii-only galaxies), and these were found to have
red colour, similar to those in our subsample.
 
In fact, while we find ellipticals to represent only about 10\% of \ea \ and \eb
\ galaxies, the occurrence of this morphological type among the
\ec \ is higher than 20\%, and it is mainly populated by the most
luminous ones. 

The occurrence of the \Oii \ line as well as other emission lines
in local elliptical galaxies, is an already known phenomenon 
\citep[see e.g.][]{caldwell84,phillips86}. These are either passive 
galaxies that are showing a residual, very low star formation level, 
or LINERS. A number of mechanisms other than star formation can 
be responsible for the production of the ionizing flux needed for the 
formation of forbidden lines: the presence of 
blue and evolved stellar populations, such as Horizontal Branch (HB) or
post--asymptotic giant branch (post--AGB) stars, energetic gas shock waves 
or cooling flows. Post--AGB stars in particular seem to be one of the most likely 
sources for such spectral features, as a recent work by \cite{singh13} 
demonstrates, especially for massive galaxies, with very low level 
of star formation activity as those in our bright \ec \ sample.

This issue is, of course, much easier do deal with for us as far as 
type-1 AGNs are concerned, as 
they are easily recognized by our spectral fitting routine. Given their 
extremely blue continuum and the presence of very
broad emission lines, which are not compatible with the typical spectral 
features of stars, in such cases 
it is not possible to obtain an acceptable fit by means of SSP spectra. 
Like this, a high $\chi^2$ value is returned, so that it is possible to 
easily recognize these objects with a targeted visual inspection.

Type-2 AGNs and LINERS have instead been identified in our
spectra by fitting the underlying stellar continuum and investigating
lines ratio diagnostic diagrams, as will be described in detail in
Marziani et al. (in prep.). However, only 0.8\% of the
galaxies in our cluster sample turn out to be AGNs (either Seyfert-1 
or Seyfert-2), therefore they are a negligible component of our 
spectroscopic sample. 

In view of these considerations, we argue that part of the  \ec--classified
objects, might very likely be constituted by galaxies with characteristics
very similar to the passive class \k \ (see Sect.\ref{sec:k}), but hosting
either some very low-energy LINER activity, or being influenced by
one or more gas ionizing mechanisms as outlined above.

The location of a galaxy with respect to its companions and 
neighbours in a cluster, the so--called
local density, plays an important role in shaping and defining some
of its characteristics. The values of the projected local density that 
we will use are calculated from a circular area enclosing the first 10 
most nearby galaxies, also accounting for background counts as 
expected from \cite{berta06}, and corrected for border effect
\citep[see][for details]{vulcani12}. The local
density is expressed as the logarithm of the number of galaxies with
$M_V\leq -19.5$ per Mpc$^{-2}$.

\begin{figure}
\centering
\includegraphics[height=0.50\textwidth]{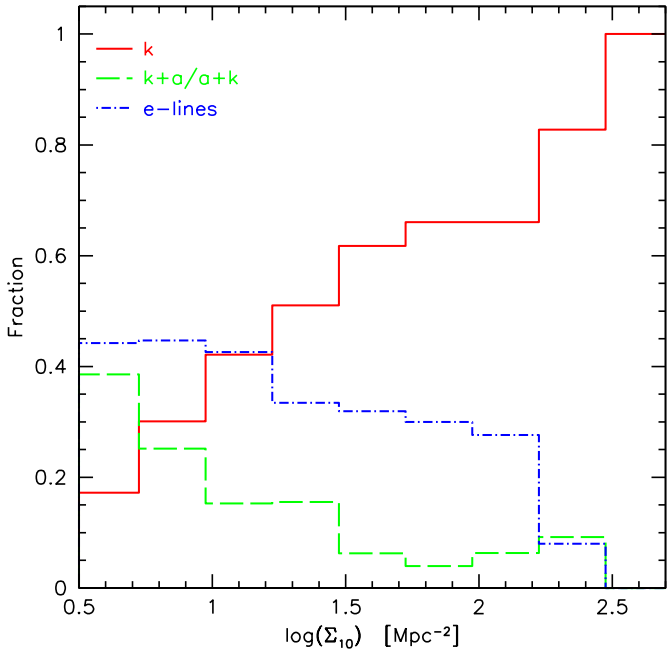}
\caption{Percentage of the passive, post--starburst and 
emission--lines types as a function of the (projected) local density.}
\label{fig:ld_spec}
\end{figure}

In Fig.\ref{fig:ld_spec} we show the fraction of galaxies of the 
different spectral types, grouped into the three broad classes 
of passive (\k), emission--lines [\ea, \eb \ and \ec] and 
post--starburst (\ka \ and \ak), at different values of the local 
density, displayed in bins of 0.25 dex. Emission--line galaxies 
dominate the regions at the lowest densities, but their relative
number monotonically decreases as the local density increases,
and they are basically not found at the highest values. 

Keeping the three emission--line classes separate slightly increases
the noise in this plot, and the interpretation does not significantly
change. \ea \ and \eb \  share a similar distribution with respect to the
local density, and they are both extremely rare at densities above 
$\sim 60$ Mpc$^{-2}$ ($\log[1.75]$): no more than 13\% of galaxies 
in these two classes are found beyond those densities. A very similar
trend is found for \ec's, whose distribution extends, instead, towards
slightly higher density values, being the main responsible for the
tail at high values observed in Fig.\ref{fig:ld_spec}. The three 
emission--line classes are distributed in a very similar 
manner compared to the two post--starburst classes, all of them 
displaying a peak between $\log(1.25)$ and $\log(1.50)$
 Mpc$^{-2}$  in their distribution. 

\subsection{The properties of the Post--Starburst population}\label{sec:poststarburst}
The \ka \ and \ak \ spectral classes, often referred to as  
post--starburst or post--star forming, make up about the 11\% 
of the local cluster population at magnitudes brighter than $-18$
($\sim 10$\% and $\sim 1$\% for the
\ka \ and \ak, respectively). This number is in strong contrast 
with the value found from the catalog of local cluster of \cite{dressler88},
but broadly consistent with a fraction of $\sim15$\% found among 
lower luminosity galaxies in 5 low-z cluster by \cite{caldwell97}. 
The reason for the striking difference with the sample of \cite{dressler88}
is probably due to the difference in the magnitude limit of the two 
surveys.

As described in Sect.~\ref{sec:morph}, when viewed as two separate 
classes, \ka \ galaxies tend to have
morphological characteristics very similar to passive objects, with a slight 
preference towards later morphological types, while
\ak \ are more similar to the emission line types in this respect. This would
fit the picture in which \ak \ galaxies are the direct descendants of 
emission lines objects, 
especially star-bursting such as the \eb \ types, gradually arising some 
$10^7$ years after the star formation has stopped, when the most massive
stars, providing the UV radiation necessary for the production of lines 
in emission, have already died. Furthermore, the fact that no \ka \ and 
\ak \ are classified as Sd/Irregular, argues against 
merging as the main mechanism to produce such spectra, and is in
the line of the findings of \cite{dressler99} in clusters at high-z: other 
processes must be the main responsible for the presence of such
galaxies in clusters. \cite{ma08} reached a similar conclusion by 
studying a massive cluster at z=0.55, being able, by analysing the 
positions and velocities they display in the cluster, to rule out mergers as
the dominant mechanism driving their creation and evolution, at least 
in clusters.

Remarkably enough, while not extremely different in the distribution of the
morphological types, \ka \ and \k \ galaxy differ substantially both with 
respect to their mass and luminosity function (see also Sect.~\ref{sec:k}); 
this is expected as an effect of downsizing, where the most massive galaxies
have already evolved into passive types.

We note that there is a significant difference in the V-band luminosity distribution 
between the two post--starburst classes (see Fig.~\ref{fig:sptype}), with
the \ka \ displaying a tail towards high luminosity, which is completely 
missing in \ak. This might be an indication that, at least part 
of the \ka \ types (the high luminosity ones), are not coming from 
\ak.
 
As the number of \ak \ is quite small when compared to all the other classes,
in the rest of our analysis we will consider these two classes as 
one, adopting the broad definition of post--starburst for both.
\begin{figure*}
\begin{tabular}{cc}
\rotatebox{270}{
\includegraphics[height=.480\textwidth]{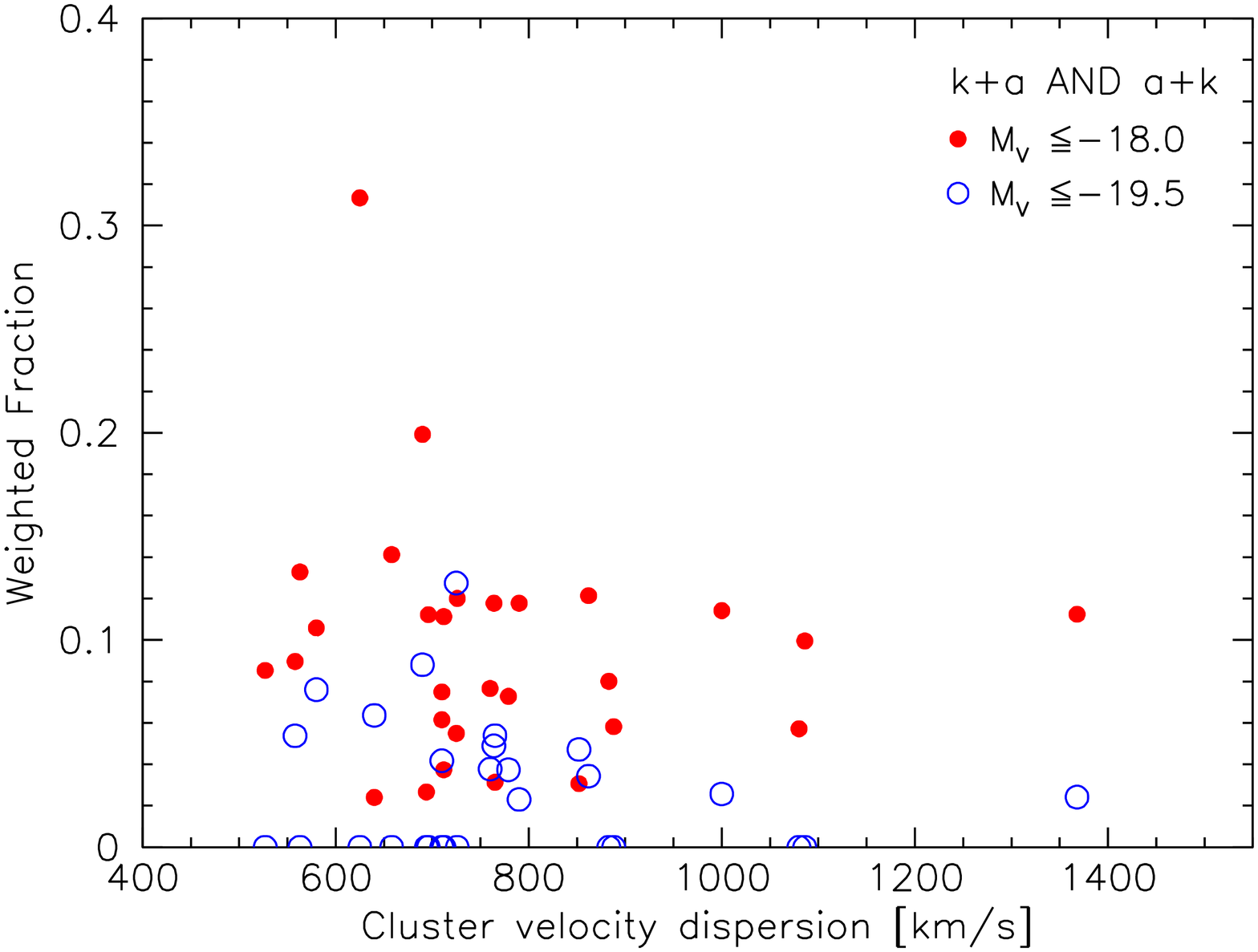}} &
\rotatebox{270}{
\includegraphics[height=.480\textwidth]{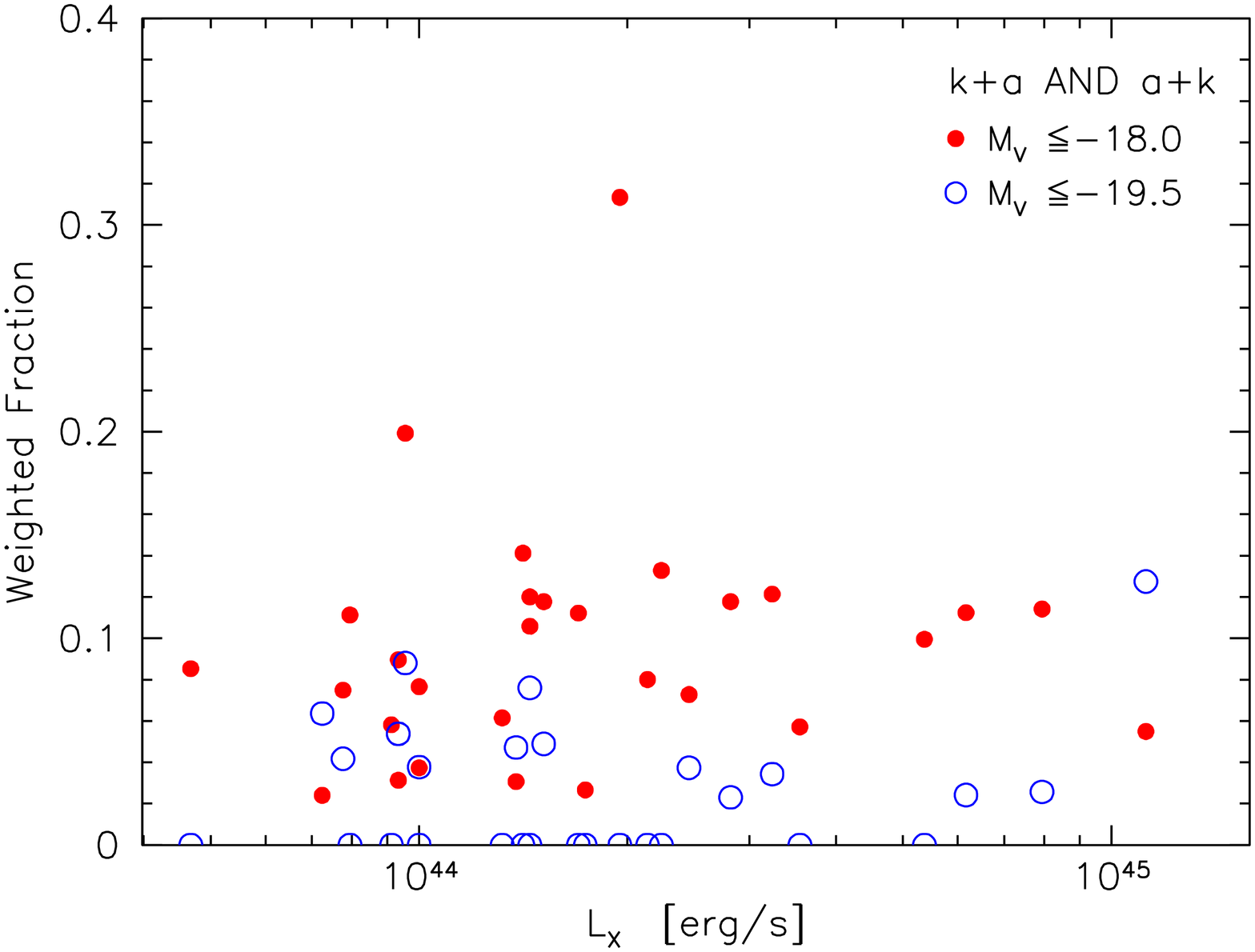}}  \\
\end{tabular}
\caption{Fraction of post--starburst galaxies,
again as a function of the two cluster mass tracers, velocity dispersion
({\it left}) and $L_X$ ({\it right}). {\it k+a} and {\it a+k} are considered as one single class. 
As done for Fig.~\ref{fig:oiifrac}, only galaxies within R$_{500}$ are 
taken into account. Points with different colours refer to different magnitude
limits and will be used later on for a comparison with high redshifts samples.}
\label{fig:psbfrac}
\end{figure*}

Similarly to what we find for emission line galaxies, we do not see any 
correlation between the fraction of post--starburst and the cluster velocity 
dispersion or X-ray luminosity (see Fig.~\ref{fig:psbfrac}): the average 
fraction, calculated over all the clusters, is $8.7\pm 5.4$. This is at odds 
with results in the distant clusters, where the incidence of such galaxies
strongly correlates with the cluster velocity dispersion \citep{poggianti09b}.

\cite{poggianti09b} define as ``quenching efficiency'' the ratio between the 
number of post--starburst and the active fraction, where ``active'' includes 
both emission line galaxies, but also the 
recently star forming classes \ka \ and \ak. We check
whether this fraction, computed from weighted counts, correlates with any
of the cluster properties. The value we have calculated by averaging
over each cluster in our sample is $20.7 \pm11.7$\%, and we see no 
significant dependency on either the X-ray luminosity or the velocity dispersion.

The post--starburst class has, in general, properties which are in 
between the passive and the star--forming galaxies. For example, 
the stellar mass distribution peaks at mass values 
which are intermediate between the \k \ and the
\ea \ and \eb \ class (see Fig.~\ref{fig:mass_sp}). The projected radial
distribution of the spectral types tells a similar story: passive
galaxies display the average tendency of gathering 
towards the cluster centre, while emission lines objects are more frequent
in the outskirts, and post--starbursts occupy a region somehow in between.
\begin{figure*}
\begin{tabular}{ll}
\includegraphics[height=.48\textwidth]{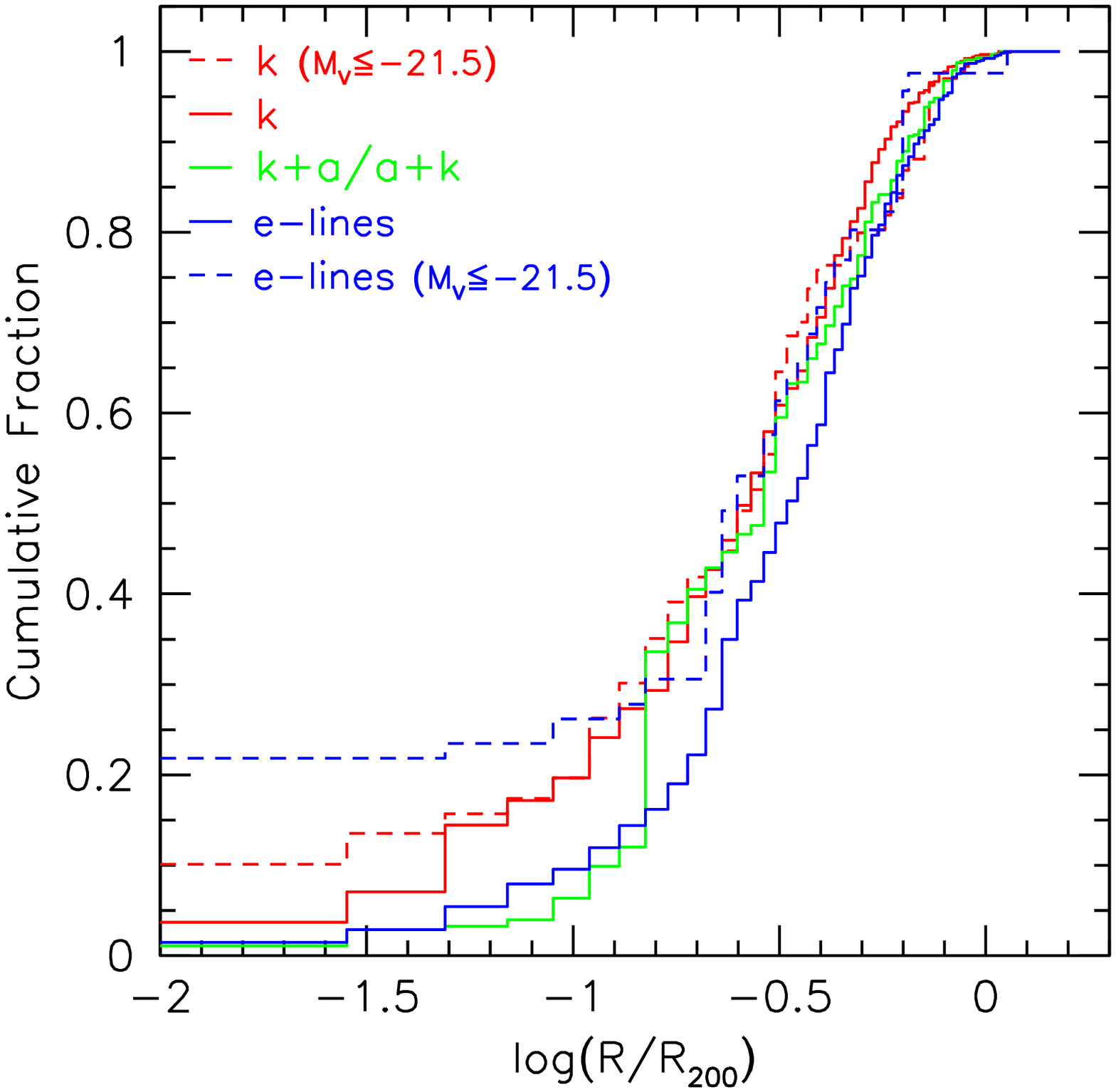} &
\includegraphics[height=.48\textwidth]{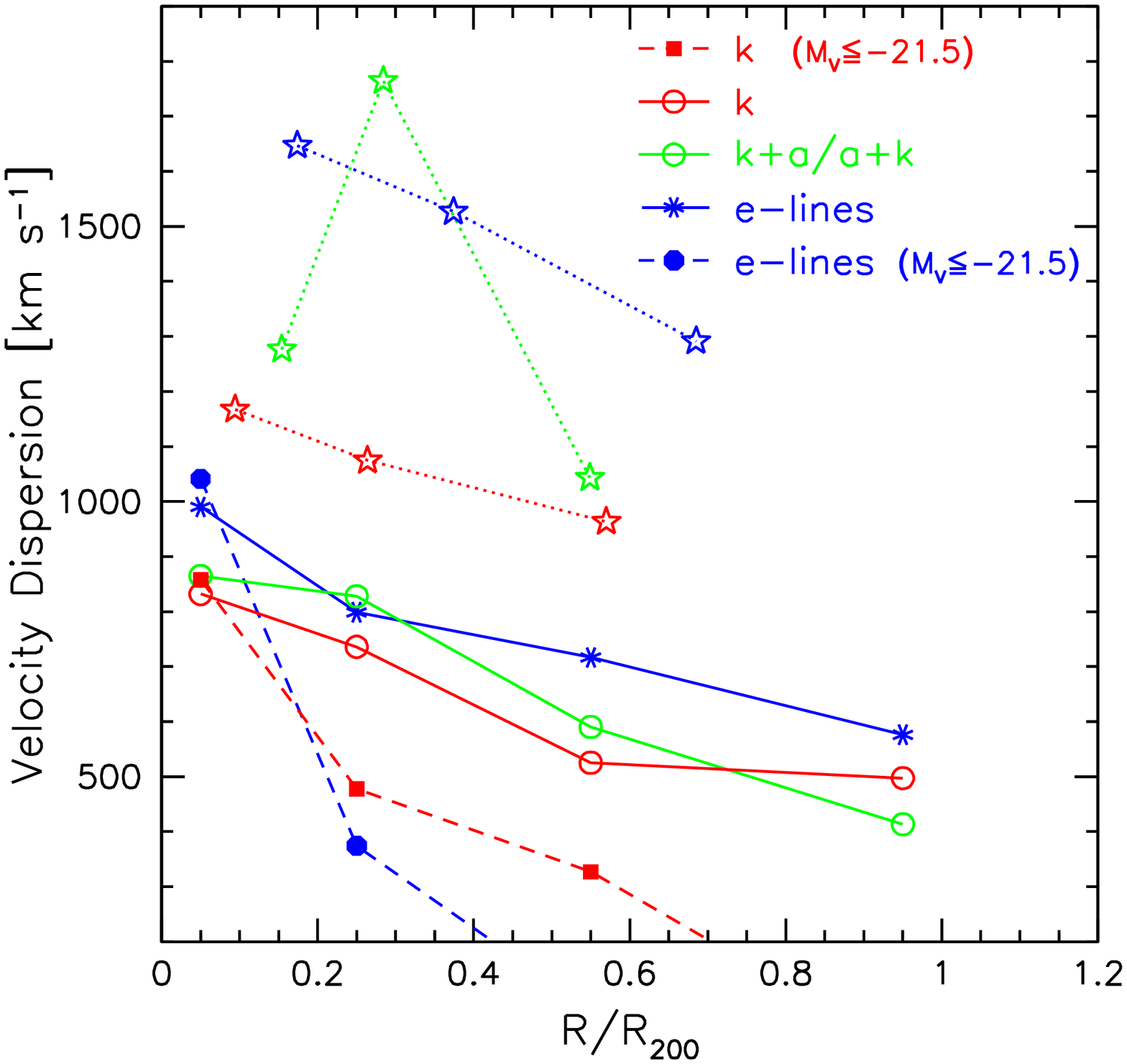} \\
\end{tabular}
\caption{On the {\it left-hand} panel: cumulative radial distribution 
of the three spectral types as defined in the text. On the {\it right-hand} 
the average velocity dispersion in bins of radial distance, for the
three main spectral classes. Void stars represent the high-z data points
from \cite{dressler99}. In all cases the distances are calculated 
from the cluster's BCG, and are expressed in terms of R$_{200}$.}
\label{fig:rad_dist}
\end{figure*}
This can be clearly seen in the left-hand panel of Fig.~\ref{fig:rad_dist}, 
where we plot the cumulative radial distribution of \wings \ spectral 
type, broadly divided in 3 classes. 

On the right-hand panel of the same figure, we show the average 
velocity dispersion in 4 bins of projected radial distances. The average
velocity dispersions of the post--starburst class have generally
intermediate values between the passive and emission line classes, as
if they were in an intermediate state of virialization between the passive and 
actively star--forming galaxies. The result we get by combining the 
information provided by  the two panels in this Figure, indicates 
ram--pressure stripping as the most effective mechanism
for the creation of a \ka/\ak \ spectrum. When a galaxy falls into a cluster, 
star formation is triggered as it reaches the cluster's virial radius \citep[e.g.][]
{byrd90,fujita98}, 
and it is then abruptly quenched once the galaxy enters deeper into the cluster
environment and its interstellar medium gets very efficiently stripped 
by the intracluster hot gas \citep{gunn72,treu03,ma08,abadi99}. The fact that we 
see post--starburst galaxies in a shell--like radial distribution, somehow
suggesting they define a transition region, might indeed be reflect the 
result of such a picture.

Finally, when considering the influence of the environs onto a galaxy 
through the local density (see Fig~\ref{fig:ld_spec}), post--starbursts quite follow the trend
of emission lines galaxies, contributing gradually less to the total population,
as the local density increases. Already at densities values larger than $\sim 30$
Mpc$^{-2}$ they make up only the 10\% of the total galactic 
population, while being almost as numerous as the emission line class
at the lowest density values. 

\subsection{Passive galaxies}\label{sec:k}
This class includes all those galaxies whose spectrum does not show 
any \Oii \ emission (see Sect.~\ref{subsec:spclass} and 
Tab.~\ref{tab:ewclass}) and weak \Hd \ absorption. 
By adopting this definition, we can consistently compare the properties
of galaxies in our sample to those at higher redshift. 
Of course, this potentially leaves out other lines such as, for example, 
the \Ha+{\sc[Nii]}, triplet which could in principle be detected even when no
\Oii \ emission is measured. We checked the
values of the \Ha \ line in the \k \ sample, and found that in only $\sim 2$\% 
(57 spectra) our method was able to detect an emission. These 57 objects
do not show any preference in the values of the stellar mass or luminosity
or other specific common properties.

As already shown, the great majority of \k \ galaxies are early types, from a
morphological perspective, around the 86\% of them being classified
as either ellipticals or S0s. This class is also dominated by higher luminosity
galaxies, and the luminosity distribution, peaking at M$_V\simeq-20.5$, 
is radically different from that of any of the other classes.
 
Similarly, the stellar mass distribution is substantially different from that of
the other 5 spectral classes: this class hosts the most massive galaxies, 
and only 20\% of them have a mass lower than $10^{10}$ M$_\odot$ in our
magnitude limited sample, in clear contrast with all the other classes where 
most of the galaxies have instead masses below that limit (see 
Fig.~\ref{fig:mass_sp}); the \ec \ galaxies are the only exception.

Following what done for the \ec \ type, we distinguish passive galaxies in two 
ranges of V-band luminosity as well, namely brighter and fainter than 
M$_V=-21.5$. As noted by \cite{biviano02} \citep[but see also][]{beisbart00}, 
the brightest early types should display the highest degree of segregation.

This is somehow confirmed for the \wings \ sample as well, as it can be seen in 
the left-hand panel of Fig.~\ref{fig:rad_dist}: the high--luminosity \k \ spectral 
type galaxies are proportionally more numerous than all other spectral types
(with the exception of the luminous emission--lines, which we have already 
discussed in Sect.~\ref{sec:emlines}), in the very central regions
of clusters, even though the difference is not as evident as the one found by
\cite{thomas06}. 
  
As for the distribution of the passive class as a function of the local density, 
Fig.~\ref{fig:ld_spec} clearly shows how galaxies in this class are the 
ones detected in the widest range of projected densities, with objects found in 
basically any kind of local environment. They make up about the 60\% of 
the galaxies at densities larger than $\sim 50$ Mpc$^{-2}$, and they are 
basically the only ones found, in our magnitude--limited sample, 
at $\Sigma_{10} \gtrsim 100$ Mpc$^{-2}$. 

\subsection{Galaxies in local clusters}\label{sec:summary}
In this section we give a summary of the properties which describe
the local population of cluster galaxies, and the picture which emerges from 
our analysis. This allows to more easily compare low and high-z 
galaxies, as we do in the following Section.

About 50\% of the galaxies in local clusters have spectra resembling those 
of passively evolving objects, while almost 40\% display emission lines and
only about 10\% show the signatures typical of objects in a 
post--starburst/post--starforming phase (see Table~\ref{tab:ewclass}). 
Passive galaxies are not only, on average, the most
massive ones, but also the most luminous, while the emission line classes 
clearly show a deficit of luminous/massive objects (see the left panel of 
Fig.~\ref{fig:sptype}). This picture very well
fits the downsizing scenario, in which the most massive galaxies are the 
first that are formed, and are already evolving in a passive fashion (i.e. following
the stellar evolution) since at least z$\sim1$ \citep[e.g.][]{depropris07,andreon08,strazzullo10}, 
while at progressively lower masses we find objects with increasingly higher 
specific star formation rates.

A remarkable exception to this picture is that of the \ec \ class, which 
contains a family of luminous (and more massive than the average) galaxies
almost resembling the high luminosity tail of the passive type. When analysed 
with a closer look, these turn out to have characteristics which are very similar 
to those of the \k's, being only different in the presence of the \Oii \ line which is 
anyway ---on average--- not very intense and it is likely produced by physical 
processes different than star formation. 

When comparing the luminosity functions of the spectral types, it is tempting to
view them as mirroring an evolutionary sequence, going from the star--forming
\eb \ and \ea, containing the least luminous galaxies, to the more ``extreme''
post starburst class \ak, whose luminosity distribution is similar to that of the
\eb \ and \ea's; containing a higher fraction of luminous object, galaxies in the
\ka \ class are characterized by a luminosity distribution in between the star-forming 
and the passive types \k \ which host, in turn, the most
luminous objects. The \ec \ are, once again, an exception 
to this scheme, both because their classification might be influenced by 
the nature of the \Oii \ line production, and because they probably contain 
the field galaxies which are being accreted by the clusters and will eventually
end up ``feeding'' the high luminosity tail of the \ka \ family.

Contrary to the findings of \cite{poggianti09}, who found the spirals fraction 
in \wings \ clusters to correlate with the X--ray luminosity, we note how neither 
the fraction of emission--lines galaxies nor that of post--starbursts, depends on the global 
properties of the cluster, specifically on the clusters' velocity dispersion or 
their X--ray luminosity, as it can be noted from Fig.~\ref{fig:oiifrac} and
Fig.~\ref{fig:psbfrac}. What drives the spectral characteristics is instead 
the local environment, as the dependence of the spectral type on the 
projected local density demonstrates (see Fig~\ref{fig:ld_spec}): galaxies
in dense environment tend to have their star formation quenched. 

\section{A comparison with the high-redshift clusters population}\label{sec:highz}
In this section we attempt a comparison of some of the properties of the 
local cluster galaxies with those at higher redshift. This way, we can 
trace the evolution of these properties as a function of the cosmic time, 
to shed some light on the physical mechanisms that drive of the physical and structural 
changes of galaxies in dense environments.

Two surveys provide samples which turn ideal to perform such a comparison: 
one was collected by the 
MORPHS\footnote{\texttt{http://www.astro.dur.ac.uk/$\sim$irs/morphs.html}} 
collaboration, which observed galaxies in clusters 
at redshifts in the $0.37-0.56$ range; the second is the ESO Distant Cluster 
Survey\footnote{\texttt{http://www.mpa-garching.mpg.de/galform/ediscs/}} 
\citep[EDisCS;][]{white05}, targeting instead cluster galaxies with 
redshifts between 0.4 and $\sim 1$. 

To properly compare the high-z samples and \wings, minimizing the effect of
possible biases, the magnitude limits of these surveys must be matched and, 
furthermore, these must be also corrected to account for evolution effects.
That is, the changes occurring in the mass--to--light ratios as the stellar 
populations in galaxies gets progressively older at different cosmic ages,
should be taken into account. If we assume that
stellar evolution is the only mechanism acting in the luminosity
change, and if we consider a common redshift of formation, we can exploit stellar
populations models to calculate the change in luminosity from redshift
0.55 to redshift 0.04 which represent the most extreme difference
between \wings \ and MORPHS. Assuming that the oldest stars are as old as
the Universe at a given redshift, we compare the V band luminosity of
simple stellar populations with an age as close as possible to the age
of the Universe at $z=0.04$ and $z=0.55$, that is $T_U\simeq 12.9$ 
and $T_U \simeq 8.08$ Gyr, respectively. The difference is, assuming a 
solar metallicity and a \cite{salpeter55} IMF, of $\sim 0.3$ magnitudes.

In their analysis 
of EDisCS galaxies, \cite{poggianti09b} use a cut in absolute V-band 
magnitude at $M_V= -20.1$ at $z=0.4$, which corresponds to M$_V=-19.8$ 
for \wings, while in the work by \cite{dressler99} using MORPHS data, 
the completeness limit is $M_V=-19+5\cdot\log_{10}h$, corresponding to
about $M_V=-19.5$ for \wings. We will use this latter value to compare
the high and low redshift galaxies, but we have verified that using -19.8 
does not change our results.

\subsection{Occurrence of the spectral classes}
We find that post--starburst galaxies in local clusters are less abundant
with respect to their analogous at high redshift. 
To our M$_V < -18$ magnitude limit we have classified 
about the 11\% of the local cluster population as post--starburst 
(i.e. either a \ka \ or a \ak).
At brighter magnitudes, matching the limits of the two aforementioned 
high redshift studies, the post-starburst fraction is 18\% at high-z 
\citep[as calculated for MORPHS galaxies by][]{dressler99}, while it
lowers down to 4.6 \% in the \wings \ clusters (if we assume a brighter
limit, at M$_V < -19.8$ the fraction further reduces to 4.4\%). These numbers
are spot on those predicted by \cite{poggianti09b} for $z\simeq 0$ 
clusters, under the hypothesis that the star formation quenching 
efficiency is similar between local and distant clusters. 
In fact, assuming post--starburst galaxies are formed after star 
formation is stopped in infalling star forming systems, and adopting 
the same value of the quenching efficiency --as defined in 
Sect.~\ref{sec:poststarburst}-- of 23\%, calculated at high redshift,
and a star forming fraction of 20\% for the local clusters 
\citep[taken from the study of][exploiting \sdss \ data]{poggianti06}, 
the expected fraction of local post--starburst would be 4.6\%. 

Indeed, if we calculate the quenching efficiency for galaxies in the
\wings \ sample, we find a value of $22.0 \pm 0.6$\%, which reduces
to $15.5\pm1.0$ if a magnitude limit of -19.5 is considered. In their
work, using both EDisCS and MORPHS data, \cite{poggianti09b} 
found a dependency of the quenching efficiency to the cluster
velocity dispersion (see their Fig.5), which is particularly evident
for the highest velocity dispersion values. This dependence is 
instead absent in \wings \ clusters (see discussion below). Hence,
the aforementioned $\sim 15$\% value is mainly to be intended for
clusters with velocity dispersions below $\sim 1000$ km/s and is, 
within this limit, in fairly good agreement with the high-z result.

Unlike their higher redshift progenitor, we find at most a weak dependence
of the relative incidence of post--starburst galaxies on total
cluster mass as traced by the velocity dispersion and the X-ray
luminosity. Looking at Fig.~\ref{fig:psbfrac} it is especially 
evident that there is a flat trend in the occurrence of these class 
as a function of the cluster X-ray luminosity, while it seems instead 
that a weak anticorrelation is present when the same quantity is
considered as a function of the velocity dispersion. This is 
compatible with the interpretation that the cluster mass at low-z
has little or no effect in producing this particular spectral class, 
in particular at low luminosities where we find most of them.

At the high redshift magnitude limit, the \wings \ sample is dominated
by the passive \k \ type, containing 2/3 of the whole population, a
significantly higher fraction with respect to the 47\% found in
\cite{dressler99}. Meaningful differences are also found for the 
emission line galaxies: the fraction of \ea \ reduces from 11\% to
a local value of  2\% and, at these bright absolute magnitudes, the \eb \
class almost disappears declining from a 16\% to less than 1\%,
clearly revealing the well known Butcher--Oemler effect. Instead, the 
continuous star--forming like spectrum \ec \ experiences an 
increase when going from high to low redshifts, moving from a 
bare 5\% to 27\%, locally. While part of this increase might come
from high--z \eb \ galaxies progressively slowing down their
star forming activity, this is barely enough to make up the
local \ec \ fraction. Two effects might explain this high number
of local luminous \ec \ we measure, in comparison to high redshift
clusters. The first is of methodological
nature, and might arise from the lower detection limit that we
have assumed for the \Oii \ lines -- EW$=-2$ \AA \ vs $-5$--, 
which discriminates between
the \ec \ and \k \ type. If this was the reason, we would end up 
having a even larger fraction of local \k \ galaxies.
{Indeed, if we assign a \k \ classification to ALL those \ec \ 
galaxies with EW(\Oii)$> -5$ \AA, their fraction would considerably 
reduce, moving to a much lower 14\%, while passive \k--type
galaxies would become the dominant galaxies containing almost 
80\% of the objects. Still, this is far from the 5\% fraction observed
at high-z. 

In fact, we could be observing a second, more 
phenomenological effect, as the different \ec \
fraction might reflect the way star formation proceeds in field galaxies
that  have been recently captured by the cluster, at a lower
and steadier pace at lower redshift as opposed to a higher
and bursty-like activity at high redshift. 

Finally, we note a drastic decrease in the incidence of \ea \ galaxies 
(from 11\% to 2\%) as well. These are unlikely to have turned
into the \ka \ or \ak \ we observe in \wings \ clusters, as long as 
passive evolution only is considered (i.e. no merger or gas accretion
events). In fact, the time scale during which the spectral
features typical of A-stars dominate is around $\sim 1$ Gyr, much
shorter than the 5 Gyr period separating the redshift range we are 
investigating, and after which the Balmer lines become progressively 
fainter and, if star formation has ceased, such galaxies would 
eventually end up showing a \k \ spectrum.

Another quite striking aspect differentiating \wings \ and MORPHS 
galaxies concerns the morphological properties. Generally 
speaking, when compared to the high-z sample, we note the 
absence of irregular galaxies among the local \k, \ka \ and \ak \ 
spectral type, and a much lower incidence of the later 
morphological classes. In fact, in the distant sample,
most post--starbursts are classified as late-type, with only 
$\sim 27$\% of the objects being assigned an ellipticals or S0 
morphology \citep[but see also][for a similar results on a study at 
comparable redshift]{tran03}. This number compares to a much 
higher fraction ($\sim 70$\%) of low redshift galaxies which have 
instead a morphological early type (E or S0) classification. 

\subsection{The Luminosity distributions}
\cite{dressler99} presented a comparison of the V-band luminosity functions 
for 6 spectral types, whose definition matches the one we have adopted 
for this work. In Fig.~\ref{fig:Vcomp} we plot those same luminosity 
distributions, which we have corrected as explained above to account for passive 
luminosity evolution, together with the local ones. Having in mind that the high-z
sample is to be considered complete for M$_V \lesssim -19.5$, thus limiting
the analysis to the brightest galaxies, we can consistently compare the two 
samples. 
\begin{figure}
\includegraphics[height=.65\textwidth]{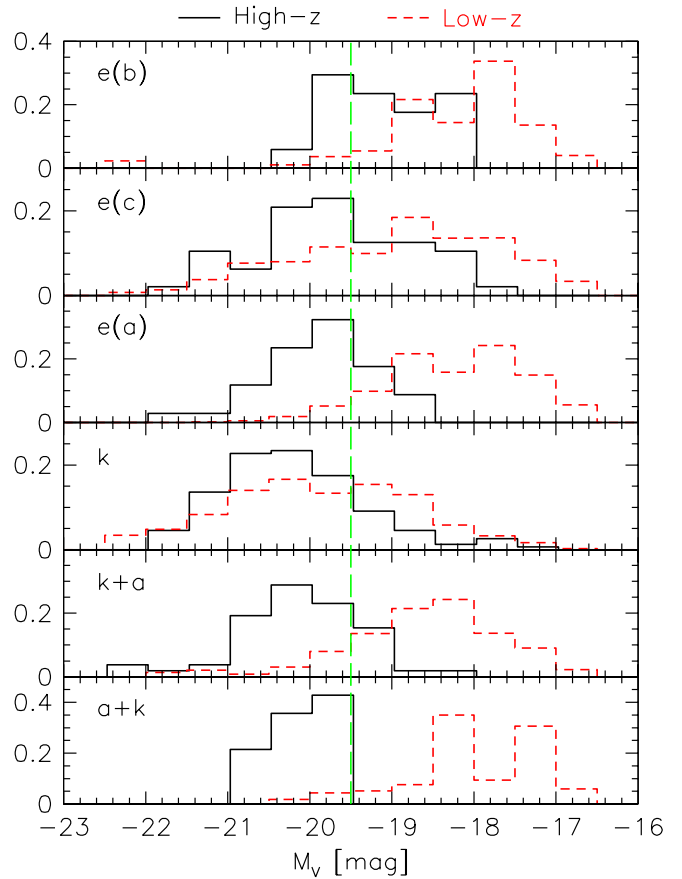} 
\caption{Comparison of the high and low redshift V-band
luminosity distribution for the 6 spectral types. High--z data
have been corrected for passive evolution so to match the 
local clusters data (see text for details). The vertical dashed
line marks the M$_V=-19.5$ limit.}
\label{fig:Vcomp}
\end{figure}

Galaxies with the strongest emission lines --\eb-- are lacking a significant
presence of objects more luminous than M$_V < -20$ throughout all
the redshift range here considered. From the uppermost panel of 
Fig.~\ref{fig:Vcomp}, it is tempting to claim that higher redshift 
star bursting
galaxies are in general more luminous compared to their most nearby
counterparts, even though incompleteness at high redshift might be
driving, at least partially, this comparison. Very luminous star bursting 
galaxies are anyway not observed in local clusters, but nothing more can 
be said on their luminosity distribution based on these datasets. 

Taking magnitude limits into account, the luminosity of 
\ec \ galaxies turns out to be very similar at these redshifts,
while the \ea \  types mostly differ due to the presence of a significantly 
brighter population in the MORPHS sample, which is 
compatible with the luminosity distribution of the low-redshift \ka \ class. It 
seems anyway unlikely that the high redshift \ea \ can be considered as 
the progenitors of the local \ka, as the elapsed time since z$\sim 0.5$ to 
now is a factor of 3-4 higher than the life of the A stars features, unless
they host a significant amount of obscured and prolonged star formation 
which will eventually leave its \ka/\ak \  signature in local clusters.

The luminosity functions of \k \ galaxies at low and high redshift are 
quite similar. Interestingly, above M$_V=-22$, the local population 
contains a slightly higher number of brighter objects. This could either be
due to a somehow incomplete sampling of the high--redshift population 
(the MORPHS sample only contains 10 clusters, as opposed to the 
$\sim30$ in \wings), or it could be a real evolution effect, for which these 
passive galaxies would have undergone dry-merger events, having 
left their spectral features unchanged, while only increasing their luminosity.

A significant difference can also be observed in the post--starburst class 
galaxies, whose high-z sample displays a) a slightly more luminous tail,
b) a higher number of galaxies with higher luminosities, and c) a significantly 
different peak magnitude, $\sim -20.2$ vs $\sim -18.5$ (even though the latter 
value may be limited by incompleteness) at high and low z, respectively. Once 
again, this fits a downsizing galaxy evolution scenario, in which the most
active galaxies are those at progressively lower mass (luminosity) as 
one moves from high to the low redshift Universe.

\subsection{Spectral properties and the position in the cluster}
Finally, we compare the radial distribution of the spectral types for the
different surveys. We consider only 3 main spectroscopic classes:
emission line galaxies, passive galaxies, including only the spectroscopic type {\it k}
and the \ka \ and \ak. 

In the left--hand panel of Fig.\ref{fig:rad_dist}, we plot the cumulative radial distribution of
the three spectral types defined above. At low redshift, the \k \ type galaxies, 
and in particular the brightest ones (M$_V \leq -21.5$), are the most 
centrally concentrated, while emission-line galaxies tend to prefer the
clusters' outskirts, and the post--starburst population generally sits at 
an intermediate region between the two.  50\% 
of \k \ galaxies are found within a projected radial distance of
 $\sim 0.25\cdot$R$_{200}$, while the same percentage for the emission lines
 ones is reached at  $\sim 0.40\cdot$R$_{200}$, showing the tendency to avoid 
 the innermost regions of the clusters, which are dominated by the early-type,
 passive population. 

Although this analysis inevitably suffers from projection
effects, it still provides some clue to how the characteristics of the
galaxies are, on average, shaped by the environment. 
These trends are similar to those found in distant clusters
by both \cite{dressler99} and \cite{poggianti09}, even though 
the differences between the three aforementioned classes seem to be
less evident for our sample. If, at low redshift, we limit our comparison
to the sample matching the distant clusters absolute magnitude, the situation
does not change for emission line and passive galaxies, even though the
radial distribution for post--starbursts becomes noise--dominated, being
hampered by the very low number of objects at this magnitude limit.

We compare also the radial distribution of the velocity dispersions for the 
spectral classes, which we have shown in the right panel of Fig.~\ref{fig:rad_dist} 
for \wings \ galaxies, between the high (MORPHS) and low z clusters. 
In general, we find that the values measured in \wings \ clusters are up to
$\sim50$\% lower with respect to the 
higher redshift counterparts, most likely due to the achievement of a more advanced 
state of virialization of the local clusters. Apart from this, the average trend
which sees the early types having the lowest velocity dispersion at any radius,
and the late types displaying a similar trend but at higher velocities, is
also seen in local clusters, even though at a much higher dispersion. Furthermore, 
the decline of the velocity as a function of the projected radius, is quite similar
for the the two samples, in all the spectral classes.

What is significantly different in local clusters is the average difference, in velocities,
in particular between passive and emission--lines galaxies, which are instead
clearly distinct at high-z. Overall, the velocity dispersions between 
the different spectral types at low redshift is much more similar, with an average
ratio $\sigma_e/\sigma_k = 1.22 \pm 0.11$, only marginally consistent with
the $1.40 \pm 0.16$ ratio of the MORPHS sample.

\section{Summary and conclusions}\label{sec:conclusions}
In this paper we have presented a catalog of equivalent widths of the most 
prominent spectral lines in the $\sim 3700$ to $6600$ \AA \ range, automatically 
measured in medium resolution spectra from the \wings \ spectroscopic 
survey. We use these measures to classify galaxies based on the intensity and 
presence/absence of the \Hd \ and \Oii \ lines, and we focus our
analysis on those objects which are spectroscopically confirmed cluster 
members. This classification
reflects the stellar content of a galaxy, and is widely used as a mean to
derive clues on the star formation history both locally and at higher 
redshifts.

Together with other physical parameters and characteristics of the \wings \ galaxies
sample derived in previous works such as morphology \citep{fasano12}, 
stellar mass and absolute magnitude \citep{fritz11}, we investigate the 
properties of the different spectral types, trying to understand how the 
characteristics of the stellar populations vary as a function of other 
galactic properties and of both local and global environment, in local clusters. 

Having used a classification scheme which is adopted by several other
studies at higher redshifts, we can then compare the characteristics of local
cluster galaxies to their more distant progenitors.

We can summarize our results as follows:
\begin{enumerate}
\item the local cluster population at magnitudes
$M_V \leq -18$ is dominated by the passive \k \ spectral type, 
followed by \ec \ galaxies, even though possibly contaminated  --especially 
at high luminosity (or stellar mass)-- by emission line from non-stellar origin. 
About 11\% of galaxies display a post--starburst spectrum;
\item \k \ galaxies are not only the most luminous ones, but they also display 
the broader range of V-band luminosities. The \ec's \ have a similar spread, but
a significantly lower ($\sim 1.5$ magnitudes) peak luminosity;
\item \k \ galaxies are also the slowest in clusters (see Fig.\ref{fig:rad_dist},
right--hand panel). This is consistent with early types being on average
more virialized with respect to late types, and with the findings that they 
were already in place since z$\simeq 0.6$ \citep[see, e.g,][]{smail97};
\item there is a broad correlation between the morphological and 
the spectroscopical classification: more than 80\% of the \k \ and \ka \ spectra are  
ellipticals or S0s, while the majority of the recently star--forming galaxies
[\ea, \eb \ and \ak \ ] are spirals. The \ec \ class is almost equally populated
by early (E+S0) and late (Sp) morphological types. Yet, the fact that most of
the spectral types display a full range of morphologies, suggests that the 
evolution of these two properties is, at least partially, decoupled;
\item no significant correlation is found between the relative number of 
emission--line galaxies and the cluster general properties, such as their 
X-ray luminosities or velocity dispersion;
\item the local density, instead, clearly plays a role in the spectral type 
segregation: a great fraction of passive galaxies is found at the largest 
values of projected density, while other spectral types are instead more 
common in less dense environment;
\item there is a fraction of the \ec \ population showing substantial 
differences with respect to the other emission line classes: they display
properties more similar to the passive type. We argue that, in
this particular case, part of them might be \k \ type galaxies where the \Oii \ line 
is not related to star formation, but has instead a different origin;
\item we confirm the lack of local high luminosity post--starbursts, 
strengthening the idea that these galaxies have undergone a significant 
evolution. The luminosity function is also significantly different with respect
to the one at high redshift;
\item local post--starbursts display the tendency to prefer earlier morphological
types, compared to the high redshift ones (i.e. no irregular \ka \ and \ak \ are
found in the \wings \ sample); 
\item the star formation quenching efficiency is similar between high and low 
redshift, as long as clusters with velocity dispersions less than $\sim1000$ km/s are
considered, with some differences possibly due to downsizing effects. This 
strongly suggests RAM-pressure stripping as the main mechanism for 
halting star formation in infalling galaxies, across the z$\sim 0-1$ redshift range.
\end{enumerate}

Finally, we briefly describe the EW catalog (see an example of the typical quantities
and format in Appendix \ref{sec:catalog}), which is made publicly available
through both the CDS an the Virtual Observatory. More details on this catalog
and, in general, on \wings \ and its products can be found in Moretti et al. 
(subm.).

\noindent
\vspace{0.75cm} \par\noindent
{\bf ACKNOWLEDGMENTS} \par

\noindent 
This work made use of Virtual Observatory tools, namely TOPCAT (http://www.star.bris.ac.uk/~mbt/topcat/).\\
Facilities: Anglo Australian Telescope (3.9 m- AAT); William Herschel Telescope (4.2 m- WHT).\\
Funding for the SDSS and SDSS-II has been provided by the Alfred P. Sloan Foundation, 
the Participating Institutions, the National Science Foundation, the U.S. Department of Energy, 
the National Aeronautics and Space Administration, the Japanese Monbukagakusho, the Max 
Planck Society, and the Higher Education Funding Council for England. The SDSS Web 
Site is \texttt{http://www.sdss.org/}.\\
We are grateful to the anonymous referee, whose comments and remarks 
helped us to improve the readability of this work.


\begin{appendix}
\section{The Equivalent Widths catalog}\label{sec:catalog}
The catalog presented in this paper contains EW values of 14 
among the most prominent emission and absorption lines in the
optical range (see Tab.\ref{tab:lines}), measured in the rest-frame 
observed spectra. Errorbars on the measures, taking into account 
both the spectral S/N and the measurement method, are also provided, with
the details of their calculation being explained in Sect.\ref{sec:ewerror}. 
These values, those of the \Oii \ and \Hd \ in particular, are used to 
derive a rough spectral classification, which is included in the catalog in
the form of a numerical flag. Emission lines galaxies are labelled with
1,2 and 3 for {\it e(a)}, {\it e(b)} and {\it e(c)}, respectively, while
the ``non-starforming'' types are flagged with 4, 5 and 6 ({\it k}, {\it k+a}
and {\it a+k} respectively). Where no classification was possible we
flag the spectrum with a 0. In Fig. \ref{fig:stat}, we report the number 
of spectra in which each line was successfully measured, for the entire 
sample, and distinguishing those that were measured in emission (red
histograms).

\begin{figure}
\centering
\rotatebox{270}{\includegraphics[height=.5\textwidth]{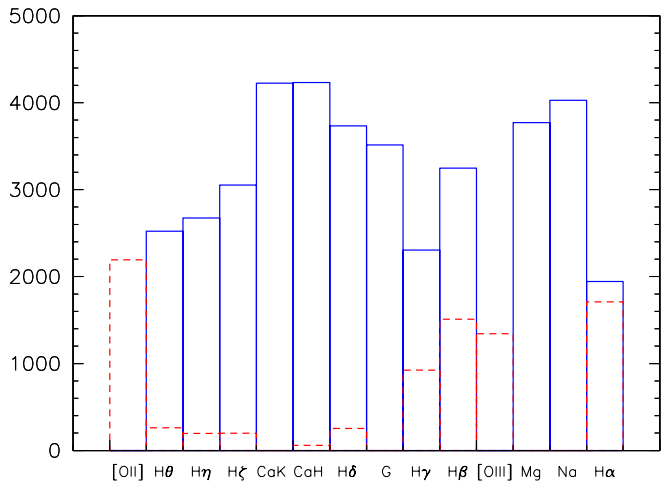}} \\
\caption{Statistics on the lines measurement. The histograms show the
number of spectra for which a given line was properly measured. Red,
dashed-line histograms refer to lines in emission (EW$<0$).}
\label{fig:stat}
\end{figure}
Non-detections and noise-dominated lines are flagged with a $99.00$,
while the value $999.00$ is used to identify those lines that lie outside 
the observed range. We also measure two classical, widely used,
indexes: $D(4000)$ \citep{bruzual83} and $D_n(4000)$
\citep{balogh99}. As already mentioned, whenever a \Oii \ and \Oiii \
EW have a value which is higher than -2 \AA, we treat this as a
non-reliable detection, which is flagged by a 0.000 value. In the
catalog we are also reporting the values of the magnitude and radial 
completeness (C($m$) and C($r$)).
Table \ref{tab:cat} shows an example of how the catalogue looks like.
Column 1 reports the \wings \ name, based on the galaxy's coordinate;
columns 2 to 15 report the equivalent width values expressed in
\AA; columns 16 to 29 report the uncertainties, following the same order
as the EW values; columns 30 and 31 contain the spectral indexes
D4000 and D$_n$4000, respectively; in column 32 we give the 
spectral class flag and in columns 33 and 34 we provide the photometric 
and geometrical completeness, respectively.

\begin{table*}
\begin{center}
\tiny
\rotatebox{90}{
\begin{tabular}{lrrrrrrrrrrrrrr} 
\hline\hline
\noalign{\smallskip}
       WINGS ID           &  \Oii   &H$\theta$&H$\eta$ & H$\zeta$& Ca{\sc k} &Ca{\sc h} &   \Hd   &CO{\sc g}&   \Hg   &   \Hb    &   \Oiii  &    Mg   & Na{\sc d}& \Ha+N[{\sc ii}]    \\ 
\hline
\noalign{\smallskip}
WINGSJ103833.76-085623.3  & -16.10  & 99.00   &  2.80  &   6.92  &   99.00   &  -0.43   &   2.48  &   3.11  &   3.96  &  -8.43   &   -9.72  &  99.00  &   99.00  &   -22.79  \\
WINGSJ103834.09-085719.2  &  -2.40  & 99.00   & 99.00  &  99.00  &    9.06   &  99.00   &   3.61  &  99.00  &  -4.87  &  -1.81   &  -44.32  &   2.22  &    2.72  &   -37.51  \\
WINGSJ103835.89-085031.5  &  -8.21  &  4.63   &  4.87  &   9.18  &    9.57   &   7.19   &   2.10  &  99.00  &  -0.66  &   5.54   &   99.00  &   1.13  &    3.94  &    99.00  \\
WINGSJ103843.03-085602.8  & -17.94  &  3.55   &  3.52  &   5.08  &    4.19   &   3.72   &  99.00  &   2.38  &  -1.89  &  -6.38   &   -3.78  &  99.00  &    1.81  &   -61.90  \\
WINGSJ103848.30-084259.9  &  -3.20  & -1.10   &  4.47  &   5.67  &    4.14   &   4.34   &   1.50  &  99.00  &   0.55  &   0.67   &    0.00  &   3.94  &    2.34  &   -18.95  \\
WINGSJ103851.34-083738.8  & -45.25  & 99.00   & 99.00  &   2.34  &    8.18   &   2.55   &  99.00  &  99.00  &  99.00  &  -1.26   &  -27.04  &   3.79  &    3.07  &   -43.93  \\
WINGSJ103857.45-085004.6  & -15.91  &  8.12   &  5.31  &   4.13  &    0.74   &   9.80   &   0.93  &  99.00  &  99.00  &  -3.08   &   -7.00  &   4.17  &   99.00  &   -47.08  \\
WINGSJ103901.91-084229.6  &  99.00  &   2.44  & 99.00  &  99.00  &    4.36   &   5.49   &   2.36  &   4.76  &  99.00  &   2.97   &   99.00  &   4.06  &    2.36  &     2.60  \\
WINGSJ103905.66-085608.3  &  99.00  &  99.00  &  0.98  &  99.00  &   12.78   &   5.29   &  99.00  &  99.00  &   2.86  &  -2.98   &    0.00  &   4.32  &    3.23  &    99.00  \\
WINGSJ103915.58-083903.6  &  -6.36  &  6.55   &  3.01  &   5.90  &    5.41   &   6.77   &  99.00  &   5.53  &   1.00  &   6.19   &   -4.10  &   1.43  &    2.89  &   999.00  \\
WINGSJ103916.50-085447.6  &  -5.28  & 99.00   &  4.10  &   7.63  &    5.08   &   6.13   &   1.90  &   3.31  &   0.76  &   0.40   &   99.00  &   3.83  &    3.19  &    -1.54  \\
WINGSJ103920.18-083428.7  &  99.00  &  99.00  &  8.46  &  99.00  &    6.82   &  99.00   &   3.13  &   2.86  &   1.22  &  -2.39   &   99.00  &   5.12  &    2.32  &    -4.58  \\
WINGSJ103920.96-084402.3  & -10.87  &  3.36   &  0.37  &   7.00  &    3.16   &   9.40   &   3.01  &   3.73  &   0.79  &   0.33   &    0.00  &   0.35  &    1.66  &   -26.99  \\
WINGSJ103921.21-082823.9  & -10.78  &  4.83   &  7.45  &  -1.28  &    2.48   &  11.36   &   3.42  &   0.83  &  99.00  &   0.99   &   99.00  &  99.00  &   99.00  &   -11.85  \\
WINGSJ103922.27-084945.3  & -17.67  &  3.87   & 99.00  &   5.38  &    4.23   &   4.18   &  10.84  &  99.00  &  99.00  &  -3.85   &  -20.88  &   4.47  &   99.00  &   -40.03  \\
WINGSJ103924.32-083214.7  & -20.25  & 99.00   & 99.00  &  99.00  &    3.93   &   4.15   &   7.66  &   3.36  &  -5.61  &  -3.33   &   99.00  &  99.00  &    0.14  &   -51.25  \\
WINGSJ103924.75-084053.8  &  99.00  & 99.00   & 99.00  &  99.00  &   11.93   &   4.07   &  99.00  &  99.00  &   0.82  &  -0.04   &   99.00  &   1.34  &    6.44  &    99.00  \\
WINGSJ103924.79-083011.3  &  -2.25  &  6.66   & 99.00  &   3.89  &    0.21   &  99.00   &   1.86  &   7.13  &   5.17  &   5.07   &   99.00  &   0.30  &    1.29  &    99.00  \\
WINGSJ103925.05-084201.6  & -18.84  & 99.00   &  1.90  &   1.66  &    1.18   &   9.53   &   5.13  &  99.00  &   6.43  &   2.65   &   -2.54  &   2.02  &    2.34  &   -41.34  \\
WINGSJ103928.74-085701.4  &  99.00  & 99.00   & 99.00  &  99.00  &   99.00   &  99.00   &   3.24  &   1.51  &   1.74  &   1.51   &  -10.92  &   1.74  &    5.28  &    99.00  \\
\noalign{\smallskip}
\noalign{\hrule}
\noalign{\smallskip}
\end{tabular} 
}
\hspace{0.5truecm}
\rotatebox{90}{
\begin{tabular}{ccccccccccccccccccc} 
\hline
\noalign{\smallskip}
\Oii$_{E}$ &H$\theta_{E}$&H$\eta_{E}$ & H$\zeta_{E}$& Ca{\sc k}$_{E}$ &Ca{\sc h}$_{E}$ & \Hd$_{E}$ &CO{\sc g}$_{E}$& \Hg$_{E}$ & \Hb$_{E}$ & \Oiii$_{E}$ & Mg$_{E}$ & Na{\sc d}$_{E}$& \Ha+N[{\sc ii}]$_{E}$ & D4000 & D$_n$4000 & cl & C($m$) &  C($r$) \\ 
\hline
\noalign{\smallskip}
 3.15  & 99.00   &  1.94  &   1.48  &   99.00   &   0.46   &   1.14  &   1.41  &   1.45  &   2.44   &    3.25  &  99.00  &   99.00  &     4.33    &  0.95  &  0.95  &   3   &  0.1807  &  0.3237   \\
 1.66  & 99.00   & 99.00  &  99.00  &    2.13   &  99.00   &   1.07  &  99.00  &   1.86  &   1.54   &   13.56  &   1.85  &    1.27  &     5.98    &  1.15  &  1.27  &   3   &  0.1807  &  0.3237   \\
 2.75  &  2.08   &  2.74  &   1.87  &    1.52   &   1.40   &   0.91  &  99.00  &   0.53  &   1.63   &   99.00  &   0.61  &    1.15  &    99.00    &  2.17  &  1.99  &   3   &  0.3636  &  0.3237   \\
 4.84  &  1.45   &  1.15  &   1.29  &    1.17   &   1.08   &  99.00  &   0.75  &   0.85  &   1.43   &    1.40  &  99.00  &    0.77  &     5.83    &  1.63  &  1.44  &   3   &  0.3659  &  0.3237   \\
 1.60  &  1.24   &  1.55  &   1.36  &    1.30   &   1.23   &   0.78  &  99.00  &   0.41  &   0.63   &    0.00  &   1.25  &    0.88  &     3.12    &  1.51  &  1.44  &   3   &  0.3659  &  0.3237   \\
15.90  & 99.00   & 99.00  &   0.75  &    1.49   &   0.87   &  99.00  &  99.00  &  99.00  &   0.89   &    4.55  &   1.23  &    0.95  &     3.90    &  1.84  &  1.49  &   2   &  0.3636  &  0.3237   \\
 3.26  &  2.90   &  1.68  &   1.01  &    0.51   &   1.57   &   0.56  &  99.00  &  99.00  &   1.02   &    2.25  &   1.80  &   99.00  &     4.15    &  1.36  &  1.24  &   3   &  0.3636  &  0.3237   \\
99.00  &   1.89  & 99.00  &  99.00  &    1.26   &   1.40   &   0.92  &   1.27  &  99.00  &   1.33   &   99.00  &   1.38  &    0.93  &     0.83    &  1.82  &  1.57  &   4   &  0.4200  &  0.3214   \\
99.00  &  99.00  & 99.00  &  99.00  &    2.95   &   1.36   &  99.00  &  99.00  &   1.29  &   1.22   &    0.00  &   1.37  &    1.05  &    99.00    &  1.78  &  1.74  &   3   &  0.3636  &  0.3237   \\
 2.51  &  1.81   &  1.25  &   1.47  &    1.30   &   1.32   &  99.00  &   2.28  &   0.59  &   2.99   &    1.79  &   0.66  &    0.97  &   999.00    &  1.68  &  1.42  &   3   &  0.1807  &  0.3947   \\
 2.72  & 99.00   &  1.53  &   1.72  &    1.25   &   1.52   &   0.88  &   1.02  &   0.49  &   0.53   &   99.00  &   1.18  &    1.02  &     0.72    &  1.91  &  1.66  &   3   &  0.3636  &  0.3237   \\
99.00  &  99.00  &  2.52  &  99.00  &    1.74   &  99.00   &   1.03  &   1.20  &   0.78  &   1.18   &   99.00  &   1.51  &    0.91  &     1.29    &  1.91  &  1.43  &   3   &  0.4200  &  0.2571   \\
 3.79  &  1.33   &  0.38  &   1.58  &    1.09   &   1.55   &   1.13  &   1.46  &   0.43  &   0.37   &    0.00  &   0.56  &    0.72  &     2.95    &  1.52  &  1.27  &   3   &  0.3636  &  0.3947   \\
 5.49  &  2.54   &  2.29  &   0.83  &    1.19   &   1.86   &   1.30  &   0.59  &  99.00  &   0.65   &   99.00  &  99.00  &   99.00  &     1.97    &  1.39  &  1.34  &   3   &  0.3636  &  0.3237   \\
 2.68  &  1.81   & 99.00  &   1.23  &    1.05   &   1.18   &   1.68  &  99.00  &  99.00  &   1.00   &    4.14  &   1.67  &   99.00  &     3.99    &  1.21  &  1.18  &   1   &  0.3636  &  0.2571   \\
 5.80  & 99.00   & 99.00  &  99.00  &    1.12   &   1.23   &   1.66  &   1.23  &   1.92  &   1.83   &   99.00  &  99.00  &    0.46  &     4.39    &  1.52  &  1.33  &   1   &  0.3636  &  0.2571   \\
99.00  & 99.00   & 99.00  &  99.00  &    3.69   &   1.65   &  99.00  &  99.00  &   0.65  &   0.46   &   99.00  &   0.89  &    2.08  &    99.00    &  3.85  &  3.04  &   3   &  0.3636  &  0.3571   \\
 1.96  &  2.83   & 99.00  &   1.18  &    0.34   &  99.00   &   0.92  &   1.85  &   1.69  &   2.19   &   99.00  &   0.55  &    0.72  &    99.00    &  1.16  &  1.09  &   3   &  0.1807  &  0.3237   \\
 3.76  & 99.00   &  1.30  &   0.81  &    0.61   &   1.69   &   1.21  &  99.00  &   1.66  &   1.26   &    0.91  &   0.94  &    0.85  &     3.56    &  1.68  &  1.42  &   1   &  0.3636  &  0.3571   \\
99.00  & 99.00   & 99.00  &  99.00  &   99.00   &  99.00   &   1.18  &   0.96  &   0.95  &   1.12   &    6.55  &   0.91  &    1.40  &    99.00    &  0.86  &  0.96  &   3   &  0.1807  &  0.3237   \\
\hline
\noalign{\smallskip}
\noalign{\hrule}
\noalign{\smallskip}
\end{tabular} 
}
\end{center}
\caption{Example of the catalog as it is made publicly available.}
\label{tab:cat}
\end{table*}

Note that -- for the spectra of the north sample -- since the
measurements of lines shortwards of $\sim 4300$ \AA \ is not reliable
in most of them, for \Oii, \Hd \ and the two
calcium lines (Ca{\sc k} and Ca{\sc h}+H$\epsilon$)
we give the values that were manually
measured, while values for the other lines -- longwards of 4300 \AA -- are
taken from the automatic measurement.

The presence of sky lines, which are not always properly subtracted from
observed spectra, might also influence both the detection and the lines'
measurement. If we consider the most prominent of such lines, i.e. those
at 5577, 5894, 6300 and 7246 \AA, we find that, as far as \wings \ cluster 
members are concerned, only the Mg and the Na lines might fall close to these
sky features, especially in the higher redshift clusters. None of the other lines
are instead affected, but we strongly suggest a visual check before using 
the values for these two lines. Similarly, a visual check should be performed in the 
spectra of higher redshift, foreground galaxies in our samples. For example, 
spectra of galaxies at z$\sim0.108$, will have both \Ha \ and the \Oiii \ lines
critically close to such sky lines.

\section{Notes on the completeness}\label{sec:completeness}
The parent catalog from which spectroscopic targets were selected has been generated adopting the following selection criteria:
\begin{enumerate}
\item V$_{tot} < 20$
\item V$_{fib} < 21.5$
\item (B-V)$_{5kpc} \lesssim 1.4$
\end{enumerate}
where V$_{fib}$ is the V-band magnitude in an aperture matching that
of the spectroscopic fiber, V$_{tot}$ is the total V magnitude and
(B-V)$_{5kpc}$ is the colour computed from a 5 kpc aperture.

The exact cut in the colour--magnitude diagram varied slightly from cluster 
to cluster due to the small differences in cluster redshift and to
minimize the level of contamination from the background. In a few
cases, the cut has purposely included a secondary red sequence, such
as for Abell 151, to be able to study also background clusters. In
order to optimize the observational setup, in a few cases galaxies at
fainter magnitudes or larger colours have been observed. These loose
selection limits were applied so as to avoid any bias in the observed
galaxy type, as is the case of a selection based on the
colour-magnitude relation only (which selects only red, early type
galaxies).

We computed magnitude and geometrical completeness from which we
define a specific weight for each galaxy in the catalog defined as:
\begin{equation}
W(m,r)_i = \frac{1}{C(m)_{i}\cdot C(r)_{i}}
\end{equation}
where $C(m)_{i}$ and $C(r)_{i}$ are the magnitude and geometrical
completeness in the opportune radial and V-band magnitude bin.

The completeness as a function of magnitude is defined as:
\begin{equation}\label{eq:mcompl}
C(m)=\frac{N_{z}}{N_{ph}}(m)
\end{equation}
where $N_z$ is the number of galaxies with measured redshifts and
$N_{ph}$ is the number of galaxies in the parent photometric catalog.
Completeness is usually a decreasing function of
the magnitude because in observations priority was given to brighter
objects. 

The success rate, that is the fraction of galaxies with
redshift determination with respect to the total number of observed
galaxies, is similarly defined as:
\begin{equation}
SR(m)=\frac{N_{z}}{N_{tg}}(m)
\end{equation}
where $N_z$ is defined as in Eq.~\ref{eq:mcompl} and $N_{tg}$ is the
number of target galaxies we actually observed. Besides that, we also
computed the radial completeness for our sample. It is known, in fact,
that fiber collision problems can lead to a variable density of
observed sources at different radii, given the fact that it is more
difficult to allocate many fibers near the crowded cluster center. 
On the other hand,
usually central parts of the clusters are privileged due to the higher
density of galaxies and observers tend to allocate as many fibers 
as possible there. Having several fibre configurations for a given cluster,
as in our case, helps mitigating the fibre collision problem even further.
The net result, in our case, is a pretty flat behavior of the radial
completeness function defined, analogously to the magnitude
completeness, as:
\begin{equation}\label{eq:rcompl}
C(r)=\frac{N_{z}}{N_{ph}}(r)
\end{equation}
with $N_z$ and $N_{ph}$ defined as in Eq.~\ref{eq:mcompl}  but for radial bins.

\section{Rejection of spectra}\label{sec:quality}
Due to the absence of calibration lines in the UV, several
spectra taken at WHT (the ``north sample''), suffer from a 
wavelength calibration
issue affecting the wavelength range shortwards of $\sim
4300$~\AA. Hence, it is often not possible to automatically recognize
and properly measure a spectral line in this wavelength range, due to
a displacement which, in the case of the \Oii \ line (which is the
bluest line that we measure), can in few extreme cases reach some
tens of \AA. Since this displacement is not only wavelength dependent,
but it may also vary from spectrum to spectrum, it is not possible to
automatically correct for this effect in a straightforward way. The
fact that it is not always possible to properly measure some of the
lines that typically characterize the stellar populations, can make
the automatic/blind spectral fitting, which was the main reason for 
having a reliable EW measure \citep{fritz07,fritz11}, meaningless 
in some cases.

To recover as much information as possible, and use as many spectra as
we could from the north sample, we proceeded as follows. After running our
fitting procedure over the entire north sample, we measured the EW
of the four most prominent UV lines, namely \Oii, the two calcium lines 
({\sc h} and {\sc k}; 3969 and 3934 \AA, respectively), and \Hd. 
This was done manually for each of these four
lines and for each spectrum displaying the calibration problem. 
Due to these issues, care must be taken when judging the quality of a 
fit by means of the $\chi^2$ value only. We recalculate a $\chi^2$ by 
only considering spectral features (i.e. lines and continuum emission)
longwards of 4300 \AA, and rejected all the spectra with a $\chi_\nu^2 > 3$.
On the remaining spectra, we checked that the values which we have 
manually measured were compatible with those of the best fit model. When 
this was the case the fit, together with the derived physical quantities (e.g.
stellar mass, star formation rate, etc.) was considered to be acceptable, 
and the spectrum considered in the analysis.

Furthermore, since after the selection procedure described above, 
some of the clusters from the north sample were
very poorly sampled only having a few galaxies that are
spectroscopically confirmed members \citep{cava09}, we decided not to
include, in all of the statistical studies that will follow in the
paper, those clusters with a number of confirmed members less than
20. This leaves us with only 7 clusters of the north sample, still 
containing more than 50\% of the usable spectra from this catalog.

\end{appendix}

\end{document}